\documentstyle[preprint,pre,aps,epsfig,amssymb]{revtex}
\begin{document}
\draft

\title{Depletion potential in hard-sphere mixtures:\\
theory and applications}

\author{R. Roth and R. Evans}
\address{H.H. Wills Physics Laboratory, University of Bristol, Bristol BS8 1TL,
United Kingdom}
\author{S. Dietrich}
\address{Fachbereich Physik, Bergische Universit\"at Wuppertal, D-42097 
Wuppertal, Germany}

\maketitle
\begin{abstract}
We present a versatile density functional approach (DFT) for calculating the
depletion potential in general fluid mixtures. In contrast to brute force DFT,
our approach requires only the equilibrium density profile of the small 
particles {\em before} the big (test) particle is inserted. For a big particle
near a planar wall or a cylinder or another fixed big particle the relevant
density profiles are functions of a single variable, which avoids the numerical
complications inherent in brute force DFT. We implement our approach for 
additive hard-sphere mixtures, comparing our results with computer simulations
for the depletion potential of a big sphere of radius $R_b$ in a sea of small 
spheres of radius $R_s$ near i) a planar hard wall and ii) another big sphere. 
In both cases our results are accurate for size ratios $s=R_s/R_b$ as small as
0.1 and for packing fractions of the small spheres $\eta_s$ as large as 0.3; 
these are the most extreme situations for which reliable simulation data are 
currently available. Our approach satisfies several consistency requirements 
and the resulting depletion potentials incorporate the correct damped 
oscillatory decay at large separations of the big particles or of the big 
particle and the wall. By investigating the depletion potential for high size 
asymmetries we assess the regime of validity of the well-known Derjaguin 
approximation for hard-sphere mixtures and argue that this fails, even for 
very small size ratios $s$, for all but the smallest values of $\eta_s$ where 
it reduces to the Asakura-Oosawa potential. We provide an accurate 
parametrization of the depletion potential in hard-sphere fluids which should 
be useful for effective Hamiltonian studies of phase behavior and colloid 
structure. Our results for the depletion potential in a binary hard-sphere 
mixture, with size ratio $s=0.0755$ chosen to mimic a recent experiment on a 
colloid-colloid mixture, are compared with the experimental data. There is 
good overall agreement, in particular for the form of the oscillations, except
at $\eta_s=0.42$, the highest value of packing fraction considered.
\end{abstract}


\narrowtext

\section{Introduction}

Two big colloidal particles immersed in a fluid of smaller colloidal particles
or non-adsorbing polymers or micelles experience an attractive depletion force
when the separation $h$ of the surfaces of the big particles is less than the
diameter of the small ones. The expulsion or depletion of the small particles
gives rise to anisotropy of the local pressure which results in the effective
attractive force between the big ones. Asakura and Oosawa and, independently, 
Vrij used excluded volume arguments to determine the effective potential 
between two big hard spheres (modeling the colloids) assuming that the small 
particles or polymers form a mutually non-interacting fluid whose centers are 
excluded from the surfaces of the colloids by a distance $R_s$ 
\cite{Asakura54}. The resulting depletion potential is attractive for 
$h<2 R_s$ and is zero for $h \geq 2 R_s$; it increases monotonically with $h$ 
from its value at contact,
$h=0$, and is proportional to $\eta_s$, the packing fraction of the small
particles [see, c.f.,  Eq.~(\ref{asabb})]. Much attention has been paid to 
depletion induced attraction within colloid science since it provides an 
important driving force for phase separation and flocculation phenomena in 
mixtures of colloids and in colloid-polymer mixtures. From a statistical 
mechanics viewpoint depletion forces are of considerable interest since they 
arise from purely entropic effects because the bare interactions between the 
particles are hard-sphere-like. Formally, it is the integrating out of the 
degrees of freedom of the small particles which gives rise to the effective 
interaction between two big ones.

Although colloid-polymer mixtures can, under favorable circumstances, be
modeled by a binary mixture of hard spheres and ideal, non-interacting polymers
(we term this the Asakura-Oosawa model), for mixtures of colloids or colloids
and micelles a more appropriate zeroth-order model is a binary hard-sphere
mixture, i.e., the small particles are not interpenetrating but experience
mutual hard-sphere repulsion. In this case it becomes a key question as to how
the depletion potential between two big hard spheres is influenced by 
interactions between the small spheres. For high packing fractions $\eta_s$, 
one might suppose that the small spheres
exhibit pronounced short-ranged correlations (layering) leading to significant
changes in the depletion potential. This would, in turn, have repercussions
for the phase behavior of the bulk mixture, making this significantly different
from that of the Asakura-Oosawa model. Such considerations have prompted 
several recent theoretical investigations of phase behavior based on an
effective one-component depletion potential description of model colloidal
mixtures \cite{Dijkstra98,Dijkstra99,Louis00}. The crucial ingredient in such
investigations is an accurate depletion potential.

Having a proper understanding of depletion potentials is not only relevant
for bulk phase behavior; it is of intrinsic interest. In recent years a
variety of experimental techniques have been developed which measure, directly
or indirectly, the depletion potential between a colloidal particle, immersed
in a sea of small colloids or polymers, and a fixed object such as a planar 
wall \cite{experiments}. Video microscopy has also been used to determine 
depletion forces for a single big colloid in a solution of small colloids 
inside a vesicle -- a system which resembles hard spheres inside a hard cavity
\cite{Dinsmore98}. Very recently Crocker et al. \cite{Crocker99} measured the
depletion potential between two big PMMA spheres immersed in a sea of small
polystyrene spheres for a range of packing fractions of the latter (see, c.f. 
Sec~\ref{sec:oscil}). At low values of $\eta_s$ the measured depletion
potential is well-described by the Asakura-Oosawa result but at higher packing
fractions the potential exhibits a repulsive barrier and for 
$\eta_s \gtrsim 0.26$ the
depletion potential is damped oscillatory with a wavelength that is of the 
order of the small particle diameter. As experiments grow in sophistication and
in resolution it is likely that further details of depletion potentials will be
revealed whose interpretation will require a reliable and versatile
theoretical approach. Such an approach should be able to tackle experimental
situations where $\eta_s$ is rather high and  to treat general `confining'
geometries. The latter include a big particle near a planar wall or in a wedge 
or cavity, as well as the case of a big particle near another, fixed big
particle. In this paper we describe such a theory for the depletion potential
based on a density functional treatment (DFT) of a fluid mixture. Our treatment
avoids the limitations of the virial expansion (in powers of $\eta_s$) and the
uncontrolled nature of the Derjaguin approximation which are inherent in
recent approaches \cite{Mao95a,Goetzelmann98} to depletion forces in 
hard-sphere mixtures. It is less cumbersome than the alternative integral
equation treatments \cite{Attard89,Attard90} and more easily adopted to
different geometries. A key feature of our treatment is that it does {\em not}
require the calculation of the total free energy of the inhomogeneous fluid or
of the local density of the small particles in contact with the big particle
\cite{Attard89,Piasecki95}. The method is much easier to implement than a
direct minimization of the free energy functional which is  numerically very 
demanding when any symmetry of the density profile of the small spheres is 
broken by the presence of the big particle.

The paper is arranged as follows: Subsection~\ref{sec:general} defines the
depletion potential in an arbitrary mixture, showing how this is related to
the one-body inhomogeneous direct correlation function of the big particles.
In Subsec.~\ref{sec:lowdens} we use this result to derive an explicit formula 
for the the depletion potential in the low-density limit, where the densities 
of all species approach zero. For the particular case of a binary hard-sphere
mixture in this limit we recover the Asakura-Oosawa result. 
Subsection~\ref{sec:asympt} describes the general asymptotic
behavior of the depletion potential for $h\to\infty$ while 
Subsec.~\ref{sec:DFT} describes the implementation of the
theory for a binary hard-sphere mixture using the DFT of Rosenfeld
\cite{Rosenfeld89}. In Section~\ref{sec:test_acc} we present several 
comparisons of our hard-sphere DFT results, for both sphere-sphere and
(planar) wall-sphere depletion potentials, with those of computer simulation
\cite{Biben96,Dickman97}. Our theory performs well for all size ratios
$s\equiv R_s/R_b$ and packing fractions $\eta_s$ for which simulation
results are available. We show that the leading-order asymptotic result for
the depletion potential provides an excellent account of the oscillations
in the calculated potential not only at longest range but also at intermediate
separations of the big spheres. Section~\ref{sec:derjag} is concerned with
assessing the regime of validity of the well-known Derjaguin approximation 
which relates
the force between two big objects to the integral of the solvation force, or
excess pressure, of the small particles confined between two planar walls 
[see, c.f., Eq.~(\ref{derj})]. We argue that this approximation is not 
reliable for the hard-sphere mixture even when the size ratio $s$ is very 
small. In Subsec.~\ref{sec:param} we describe a simple but accurate 
parametrization
of the depletion potential suitable for a big hard sphere near a planar
hard wall and for the potential between two big hard spheres. Such a
parametrized form should prove useful for effective Hamiltonian studies of
phase behavior and colloid structure \cite{Dijkstra98,Dijkstra99,Louis00}.
Subsection~\ref{sec:oscil} presents results for the depletion potential in a
binary hard-sphere mixture where the size ratio is chosen to mimic the
system considered in the experiments of Ref.~\cite{Crocker99}. We conclude
in Sec.\ref{sec:conc} with a discussion and summary of our results.

\section{The depletion potential} \label{sec:theory}

\subsection{General theory} \label{sec:general}

We consider a general mixture of $\nu$ components in which each species $i$
($i=1,\dots,\nu$), characterized by its radius $R_i$, is coupled to a reservoir
with chemical potential $\mu_i$ and is subject to an external potential 
$V_i({\bf r})$. The mixture at thermodynamic equilibrium can be described by 
the set of number density profiles $\{\rho_i({\bf r})\}$. For such a mixture we
wish to calculate the depletion potential, or the depletion force, between an 
object fixed at position ${\bf r}_1$ and a second one fixed at ${\bf r}_2$.  
Without loss of generality the position ${\bf r}_1$ of the first object is 
chosen as the origin of the coordinate system. This fixed object then exerts 
an external potential on the particles constituting the mixture. The external 
potential can represent a planar hard wall \cite{Goetzelmann99} or a fixed 
particle of the mixture, or more generally, a curved surface \cite{Roth99} or 
soft planar walls \cite{Bechinger99}. If the depletion potential between two 
particles of the mixture is to be calculated either particle can be chosen to 
act as the external potential and this point will be addressed in more detail 
in a later section.

In the following the second object is a test particle of a species denoted as
$b$. The grand potential of the mixture when the test particle is fixed at the
position ${\bf r}_b$ in the presence of the fixed object exerting the external 
potential $V_b({\bf r})$ is denoted by $\Omega_{tb}({\bf r}_b;\{\mu_i\};T)$. 
$W_t({\bf r}_b)$, the quantity of interest here, is defined as the difference 
of grand potential between a configuration in which the test particle is in 
the vicinity of the fixed object and one in which the test particle is deep in
the bulk, i.e., ${\bf r}_b \to \infty$:
\begin{equation} \label{definition}
W_t({\bf r}_b)=\Omega_{tb}({\bf r}_b;\{\mu_i\};T)-
\Omega_{tb}({\bf r}_b \to \infty;\{\mu_i\};T).
\end{equation}
In order to calculate this difference 
the test particle can be moved along any path from one configuration to the 
other. A particular path which simplifies the calculation is via the reservoir.
This path can be divided into two steps. In the first step the test particle 
is removed from the bulk at ${\bf r}_b \to \infty$ and put into the reservoir. 
In the second step the test particle is taken from the reservoir and is 
inserted back into the mixture but now at ${\bf r}_b$. The formal means to 
describe particle insertion in a general mixture is the potential distribution
theorem and we employ this in the grand ensemble \cite{Henderson83}. 

The potential distribution theorem provides an expression for the partition
function $\Xi_{tb}({\bf r}_b;\{\mu_i\};T)$ of the mixture after a test 
particle of species $b$ is inserted at position ${\bf r}_b$ in terms of the 
partition function of the mixture $\widetilde{\Xi}(\{\mu_i\};T)$ and the 
number density profile $\rho_b({\bf r})$ of species $b$ {\em before} the 
particle insertion:
\begin{equation} \label{pdt}
\Xi_{tb}({\bf r}_b;\{\mu_i\};T) = \exp \left(\beta(V_b({\bf r}_b)-\mu_b) 
\right) \Lambda_b^3~ \rho_b({\bf r}_b)~ \widetilde{\Xi}(\{\mu_i\};T),
\end{equation}
with $\beta^{-1}=k_B T$ and $\Lambda_b$ the thermal wavelength of species $b$. 
Together with a well-known result from density functional theory (DFT)
\cite{Evans79},
\begin{equation}
\Lambda_b^3~ \rho_b({\bf r}) = \exp \left(\beta(\mu_b-V_b({\bf r}))+
c_b^{(1)}({\bf r};\{\mu_i\})
\right),
\end{equation}
it follows that the one-body direct correlation function $c_b^{(1)}$ of species
$b$ can be written as
\begin{eqnarray}
c_b^{(1)}({\bf r}_b;\{\mu_i\}) & = & \ln \left( 
\Xi_{tb}({\bf r}_b;\{\mu_i\};T)/
\widetilde{\Xi}(\{\mu_i\};T) \right) \nonumber \\
& = & \beta \widetilde{\Omega}(\{\mu_i\};T) - 
\beta \Omega_{tb}({\bf r}_b;\{\mu_i\};T), 
\end{eqnarray}
i.e., $-\beta c_b^{(1)}({\bf r}_b;\{\mu_i\})$
describes the change in the grand potential of the whole system due to 
insertion of a test particle. The grand potential difference defined by 
Eq.~(\ref{definition}) can now be expressed in terms of the difference of 
one-body direct correlation functions:
\begin{equation} \label{dep}
\beta W_t({\bf r}_b) = c_b^{(1)}({\bf r}_b \to \infty;\{\mu_i\}) - 
c_b^{(1)}({\bf r}_b;\{\mu_i\}).
\end{equation}
As the potential distribution theorem, Eq.~(\ref{pdt}), is a general result, 
valid for any number of components, for arbitrary densities of all components 
and, in fact, for any inter-particle potential function  the same generality 
holds for Eq.~(\ref{dep}). No approximations have been made so far. However, 
in order to use Eq.~(\ref{dep}) to calculate $\beta W_t({\bf r})$ an explicit 
procedure that can treat a mixture must be applied. Simulations provide such a
procedure as does density functional theory. We shall consider both here.

We emphasize that the direct correlation function entering Eq.~(\ref{dep}) 
depends on the equilibrium density profiles {\em before} the test particle of 
species $b$ is inserted at position ${\bf r}_b$. This observation simplifies 
the calculation of $\beta W_t({\bf r})$ dramatically because the symmetry of 
the relevant density profiles $\{\rho_i({\bf r})\}$ is determined solely by the
symmetry of the external potentials and therefore depends only on the nature 
of the object that is fixed at the origin. If this object is a structureless
planar wall and in the absence of spontaneous symmetry breaking such as
prefreezing or crystalline layer formation the density profiles of all species
reduce to one-dimensional profiles $\{\rho_i(z)\}$ with $z$ the distance 
perpendicular to the wall. For a fixed spherical or cylindrical wall or 
particle the density profiles $\{\rho_i(r)\}$ depend only on the radial 
distance. Even if the fixed object is a wall of more general shape, so that 
there is no simple symmetry involved in the problem, calculating the density 
profiles before particle insertion is much easier than after insertion, when 
the broken symmetry due to the presence of the test particle leads to a more 
complex dependence of the profiles on the coordinates.

While Eq.~(\ref{dep}) can be evaluated for arbitrary densities of species $b$
within the present DFT approach, a particular limit in which the density of 
species $b$ goes to zero is considered now. This dilute limit is especially 
important since it arises in the context of measuring depletion forces and in 
formal procedures for deriving effective Hamiltonians for big particles by 
integrating out the degrees of freedom of the small particles. For example, if
in a binary mixture the degrees of freedom of the small particles are 
integrated out the resulting effective one-component fluid can be described by
an effective Hamiltonian containing a volume term, to which only the small 
particles contribute, a one-body term, in which a single big particle in a 
'sea' of small particles contributes, a two-body term, a three-body term  and 
so on \cite{Dijkstra98}. For highly asymmetric mixtures the most important 
contributions come from the volume and the one- and two-body terms. This 
assumption is substantiated by the results of calculations of three-body
contributions reported in Ref.~\cite{Biben96} for a size ratio $s=0.1$.
Three-body contributions also seem to be small for $s=0.2$ \cite{Melchionna00}.
Note that the two-body term describes an effective pairwise 
interaction potential between two big particles which turns out to be 
precisely the depletion potential, i.e., $\beta W_t({\bf r})$  evaluated in 
the dilute limit \cite{Dijkstra98}.

In the grand ensemble the dilute limit can be obtained by taking the limit 
in which the chemical potential of species $b$, $\mu_b \to - \infty$, with the 
chemical potentials of all other species $\{\mu_{i\not=b}\}$ kept fixed. The 
depletion potential is then given by
\begin{eqnarray} \label{dilute1}
\beta W({\bf r}) & \equiv & \lim_{\mu_b\to - \infty} \beta W_t({\bf r}) 
\nonumber \\
& = & c_b^{(1)}({\bf r}\to \infty;\{\mu_{i\not = b}\},\mu_b\to -\infty) - 
c_b^{(1)}({\bf r};\{\mu_{i\not = b}\},\mu_b\to -\infty), 
\end{eqnarray}
which contains no explicit dependence on the external potentials that are 
present, i.e., the depletion potential depends only on the {\em intrinsic} 
change of the grand potential.

Although in the dilute limit both the density profile $\rho_b({\bf r})$ and the
bulk density $\rho_b^{bulk}=\rho_b(\infty)$ of species $b$ vanish, the ratio
stays finite and the depletion potential can also be obtained from the result
\begin{equation} \label{dilute2}
\beta W({\bf r}) = -\lim_{\mu_b\to -\infty} \ln\left(\frac{\rho_b({\bf r})}
{\rho_b(\infty)}\right) - V_b({\bf r}) + V_b(\infty).
\end{equation}
which takes a more familiar form if we re-write Eq.~(\ref{dilute2}) as
$p({\bf r})/p(\infty) = \exp[-\beta (W({\bf r})+V_b({\bf r}))]$, where 
$p({\bf r})$ is the probability density of finding the particle of species $b$ 
at a position ${\bf r}$ and we assume $V_b(\infty)=0$. This route to the 
depletion potential was employed successfully in a grand canonical Monte Carlo
simulation of a big sphere in a sea of small hard spheres near a hard wall
\cite{Goetzelmann99}. It is also the route used to obtain $W({\bf r})$ from 
experiment \cite{Bechinger99,experiments,Dinsmore98,Crocker99}. 
Note that in the same limit 
$\mu_b\to - \infty$ the density profiles of all other species 
$\{\rho_{i\not= b}({\bf r})\}$ reduce to those of a $\nu-1$ component mixture.

\subsection{The low density limit} \label{sec:lowdens}

In order to implement the formal result in Eq.~(\ref{dilute1}) 
a way of determining the direct correlation function $c_b^{(1)}$ is 
required. It is convenient to adopt a DFT perspective. In density functional 
theory the intrinsic Helmholtz free energy functional can be divided into an 
ideal gas contribution plus an excess over the ideal gas contribution. While 
the former is known exactly, in general, only approximations are available for 
the excess part \cite{Evans79}. One important exception is the excess free 
energy functional for a general mixture in  the low density limit, i.e., in 
the limit of {\em all} densities going to zero. By means of a diagrammatic 
expansion it can be shown that the {\em exact} excess free energy functional 
in this limit is given by
\begin{equation} \label{lowdensfunc}
\lim_{\{\mu_i \to - \infty\}} \beta {\cal F}_{ex}[\{\rho_i\}] =
- \frac{1}{2} \sum_{i,j} \int {\mathrm d}^3 r~ \int {\mathrm d}^3 r'~ 
\rho_i({\bf r}) \rho_j({\bf r'}) f_{ij}({\bf r} - {\bf r}'),
\end{equation}
where $f_{ij}$ is the Mayer bond between a particle of species $i$ and one of 
species $j$. 

For a binary mixture in the low density limit the depletion potential acting 
on a big particle $b$ can be calculated from Eq.~(\ref{lowdensfunc}) using the 
definition of the one-body direct correlation function given within density 
functional theory,
\begin{equation} \label{onebody}
c_b^{(1)}({\bf r};\{\mu_i\}) = -\beta \frac{\delta {\cal F}_{ex}[\{\rho_i\}]}
{\delta \rho_b({\bf r})}~,
\end{equation}
and we obtain
\begin{equation}
\beta W({\bf r}) = - \int {\mathrm d}^3 r'~ (\rho_s({\bf r}') - \rho_s(\infty))
f_{bs}({\bf r}-{\bf r}'),
\end{equation}
where $s$ refers to the small particles. In the same limit the density profile 
of the small particles reduces to the density profile of an ideal gas in the 
external potential $V_s({\bf r})$, i.e., $\rho_s({\bf r}) = \rho_s(\infty)
\exp(-\beta V_s({\bf r}))$ and the depletion potential can be written as
\begin{equation} \label{lowdensity}
\beta W({\bf r}) = - \rho_s(\infty) \int {\mathrm d}^3 r'~ 
(\exp[-\beta V_s({\bf r}')]-1) f_{bs}({\bf r}-{\bf r}').
\end{equation}
This result is more familiar for the case of a binary hard-sphere mixture with 
sphere radii $R_b$ and $R_s$ where 
$f_{bs}({\bf r}-{\bf r}') = -\Theta((R_b+R_s) - |{\bf r}-{\bf r}'|)$, where 
$\Theta$ is the Heaviside function. Then Eq.~(\ref{lowdensity}) reduces to the 
well-known Asakura-Oosawa depletion potential \cite{Asakura54}. As an example 
we consider the depletion potential between two big spheres; in this case
$\exp[-\beta V_s(r)]-1 = f_{bs}(r)$ and the sphere-sphere depletion potential 
can be expressed as \cite{Goetzelmann98}
\begin{eqnarray} \label{asabb}
\beta W_{bb}^{AO}(h) & = & - \rho_s(\infty) \pi (2 R_s -h) \left\{ R_s 
[R_b+\frac{2}{3} R_s] - \frac{h}{2}[R_b + \frac{R_s}{3}] - \frac{h^2}{12} 
\right\}~~\mbox{for }~h<2 R_s
\nonumber \\
& = & 0~~\mbox{for }~h>2 R_s,
\end{eqnarray}
where $h$ is the separation between the surfaces of the big hard spheres. As a 
second example we consider a big sphere near a planar, structureless hard wall.
The depletion potential is then
\begin{eqnarray} \label{asawb}
\beta W_{wb}^{AO}(h) & = & - 2 \rho_s(\infty) \pi (2 R_s - h) \left\{ R_s 
[R_b + \frac{R_s}{3}] - \frac{h}{2}[R_b - \frac{R_s}{3}] - \frac{h^2}{6} 
\right\}
~~\mbox{for }~h<2 R_s \nonumber\\
& = & 0~~\mbox{for }~h>2 R_s,
\end{eqnarray}
where $h$ is the separation between the surface of the big hard sphere and the 
hard wall. 

It is important to recognize that Eq.~(\ref{lowdensity}) provides the exact
low density expression for the depletion potential even if the interactions 
between the species or between the wall and species $s$ are soft and possibly
contain an attractive part so that the Mayer $f$ functions cannot be
expressed in terms of the Heaviside 
function $\Theta$. In general there is also a direct interaction potential 
between two big particles, or between a single big particle and a wall, so 
that the total effective potential, after integrating out the degrees of 
freedom of the small particles, is the sum of the {\em intrinsic} contribution
-- the depletion potential -- and the direct interaction potential 
$V_b({\bf r})$, i.e., 
\begin{equation} \label{phitot}
\Phi_{tot}({\bf r}) = W({\bf r}) + V_b({\bf r}).
\end{equation}
For example the total effective potential between two big hard spheres in the 
sea of small hard spheres is $\Phi_{tot}(r) = W(r) + V_b(r)$, with $V_b(r)$ 
the hard-sphere potential between the two big ones, and it is $\Phi_{tot}(r)$ 
which constitutes the effective pair potential in the effective one-component 
Hamiltonian for the big spheres \cite{Dijkstra98}.

It is instructive to note that the functional given by the r.h.s. of
Eq.~(\ref{lowdensfunc}) generates the appropriate depletion potential for the 
original Asakura-Oosawa model \cite{Asakura54} of a mixture of colloids and 
{\em ideal}, non-interacting polymers. This model binary mixture is specified 
by $f_{cc}$, $f_{cp}$, and $f_{pp}$, the Mayer $f$ functions describing the 
pairwise interactions between two colloids, between a colloid and a polymer,
and between two polymers, respectively. $f_{pp}$ is set to zero in order to 
describe the ideal, non-interacting polymer coils. The resulting depletion 
potential is still given by Eq.~(\ref{lowdensity}) but this result now holds 
for {\em all} polymer densities $\rho_s(\infty)$ not just in the dilute limit 
$\rho_s(\infty)\to 0$, because the polymer is taken to be ideal. The total 
effective potential between two colloids is then given by Eq.~(\ref{phitot}), 
with the Asakura-Oosawa result (\ref{asabb}) for $W(r)$, which may be employed
in an effective Hamiltonian for the colloids \cite{Dijkstra99}.

\subsection{Asymptotic behavior} \label{sec:asympt}

In the previous subsection we showed that for the case of hard spheres the 
depletion potential reduces in the low density limit to the Asakura-Oosawa 
result. Examination of Eqs.~(\ref{asabb}) and (\ref{asawb}) shows that this 
potential is identically zero for separations $h$ between the spheres or 
between the sphere and the wall that are greater than 2$R_s \equiv \sigma_s$. 
Outside the low density limit of the small particles this is no longer valid. 
From the general theory of the asymptotic decay of correlations \cite{Evans94} 
it is known that for systems in which the interatomic forces are short-ranged, 
i.e., excluding power-law decay, the density profiles of {\em both} components
of a binary mixture exhibit a {\em common}
damped oscillatory form in the asymptotic regime, far from the wall or fixed 
particle, which is determined fully by the pole structure of the total pair 
correlation functions $h_{ij}(r)$ of the bulk mixture. The depletion potential
is related to the density profile of species $b$ via Eq.~(\ref{dilute2}) and 
therefore its  asymptotic behavior should be related directly to that of this
density profile. In order to understand this connection in more detail we 
first recall some arguments from Ref.\cite{Evans94}.

A bulk binary mixture consisting of small particles of density $\rho_s^{bulk}$ 
and big particles of density $\rho_b^{bulk}$ is considered. The total 
correlation functions in the bulk, $h_{ij}(r)$, with $i,j=s,b$ are related to 
the radial distribution functions $g_{ij}(r)$ via $h_{ij}(r)=g_{ij}(r)-1$ and 
to the two-body direct correlation functions $c^{(2)}_{ij}(r)$ via the 
Ornstein-Zernike relation for mixtures. In Fourier space the latter can be 
expressed as
\begin{equation} \label{totalcorr}
\hat h_{ij}(q) = \frac{\hat N_{ij}(q)}{\hat D(q)}
\end{equation}
where $\hat h_{ij}(q)$ is the 3 dimensional Fourier transform of $h_{ij}(r)$,
the numerator is given by
\begin{equation}
\begin{array}{lcl}
\hat N_{aa}(q) & = & \hat c_{aa}^{(2)}(q) + \rho_b^{bulk} 
(\hat c^{(2)}_{ab}(q)^2-\hat c^{(2)}_{aa}(q) \hat c^{(2)}_{bb}(q)), \\
\hat N_{bb}(q) & = & \hat c_{bb}^{(2)}(q) + \rho_a^{bulk} 
(\hat c^{(2)}_{ba}(q)^2-\hat c^{(2)}_{aa}(q) \hat c^{(2)}_{bb}(q)), \\
\hat N_{ab}(q) & = & \hat c^{(2)}_{ab}(q), 
\end{array}
\end{equation}
and 
\begin{equation} \label{denom}
\hat D(q) = (1-\rho_s^{bulk} \hat c^{(2)}_{ss}(q)) (1-\rho_b^{bulk} 
\hat c^{(2)}_{bb}(q)) - \rho_s^{bulk} \rho_b^{bulk} \hat c^{(2)}_{sb}(q)^2
\end{equation}
is a common denominator.
The total correlation function in real space can be obtained by taking the 
inverse Fourier transform:
\begin{equation}
r h_{ij}(r) = \frac{1}{2 \pi^2} \int \limits_0^\infty {\mathrm d} q~ q 
\sin(q r)~ \hat h_{ij}(q)
\end{equation}
which can then be evaluated by means of the residue theorem. If $q_n$ denotes 
the $n$-th pole in the upper complex q half-plane and $R_n$ the corresponding 
residue of $q \hat h_{ij}(q)$, the total correlation function can be written 
as \cite{Evans94}:
\begin{equation} \label{residue}
r h_{ij}(r) = \frac{1}{2 \pi} \sum_n e^{i q_n r} R_n.
\end{equation}
From this equation it becomes clear that the asymptotic behavior of 
$h_{ij}(r)$ is dominated by the pole or poles $q_n$ with the smallest 
imaginary part, since this gives rise to the slowest exponential decay.

For all pairs $i,j=b,s$ the poles are determined by the condition $\hat D(q)=0$
\cite{Evans94}. For a binary mixture in the dilute limit of the big particles, 
i.e., $\rho_b^{bulk}\to 0$, the general theory of the asymptotic decay 
simplifies considerably and from Eq.~(\ref{denom}) we see that the pole 
structure of all three total correlation functions can be obtained from the 
solutions of the equation
\begin{equation} \label{condition}
1-\rho_s^{bulk} \hat c^{(2)}_{ss}(q) = 0,
\end{equation}
with $\hat c^{(2)}_{ss}(q)$ referring to the fluid of pure $s$ at density 
$\rho_s^{bulk}$. In general there will be an infinite number of solutions of 
Eq.~(\ref{condition}), but only the solution $q_n \equiv q=a_1+i a_0$ with the 
smallest imaginary part $a_0$ is important for the following. The asymptotic 
behavior of the radial distribution functions $\rho_i(r)/\rho_i^{bulk}$ of 
species $i$ around a fixed particle, which can be a small or a big one, 
can be ascertained and it follows that the density profiles exhibit asymptotic 
decay of the form
\begin{equation} \label{lowdensprof}
\rho_i(r)-\rho_i^{bulk} \sim \frac{A_{pi}}{r} \exp(-a_0 r) \cos(a_1 r - 
\Theta_{pi}),~~r \to\infty,
\end{equation}
with a common characteristic inverse decay length $a_0^{-1}$ and wavelength of
oscillations $2 \pi/a_1$ for both species $i=s,b$. Remarkably, exactly the 
same inverse decay length and wavelength also characterize the asymptotic 
decay of the density profiles close to a planar wall. This is given by 
\cite{Evans94,Evans93}
\begin{equation} \label{lowdensprof2}
\rho_i(z)-\rho_i^{bulk} \sim A_{wi} \exp(-a_0 z) \cos(a_1 z - \Theta_{wi}),
~~z \to\infty,
\end{equation}
for $i=s,b$. The amplitudes $A_{pi}$ and $A_{wi}$ and the phases 
$\Theta_{pi}$ and $\Theta_{wi}$ do depend on species $i$ and whether a 
particle or wall is the source of the external potential. Note that from 
Eq.~({\ref{condition}) it follows that in the dilute limit for the big
particles $a_0$ and $a_1$ are functions of the packing fraction of the small 
particles only; thus they do not depend on the size ratio.

The asymptotic behavior of the depletion potential can now be obtained from
Eq.~(\ref{dilute2}). Assuming that the external potential acting on the big 
spheres is of finite range the depletion potential between two big spheres
has an asymptotic behavior of the form
\begin{eqnarray} \label{asymptpot1}
\beta W(r) & \sim & -\ln\left(1+ \frac{A_{pb}}{r} \exp(-a_0 r) \cos(a_1 r -
\Theta_{pb})/\rho_b^{bulk} \right) \nonumber \\
& \sim & -\frac{A_{pb}}{r} \exp(-a_0 r) \cos(a_1 r - \Theta_{pb}) /
\rho_b^{bulk},~~r \to\infty,
\end{eqnarray}
and that between a single big sphere and a planar wall takes the form
\begin{equation} \label{asymptpot2}
\beta W(z) \sim -A_{wb} \exp(-a_0 z) \cos(a_1 z - \Theta_{wb}) /\rho_b^{bulk},
~~z \to\infty.
\end{equation}
We shall see later that our DFT results for the density profiles and the 
depletion potential conform with these asymptotic results at very large 
separations and, strikingly, at {\em intermediate} separations.

\subsection{A density functional approach for hard spheres} \label{sec:DFT}

For the system of primary interest, namely the mixture of hard spheres, a
very reliable DFT exists, namely the Rosenfeld fundamental measures functional 
\cite{Rosenfeld89}. While, in principle, this functional also can treat 
generally shaped convex hard particles \cite{Rosenfeld94}, its application has
been restricted to the particular cases of hard spheres and parallel 
hard cubes \cite{Cuesta}.

In the low density limit the Rosenfeld functional reduces to the exact excess 
free energy functional of Eq.~(\ref{lowdensfunc}). For arbitrary densities it 
has the following structure:
\begin{equation} \label{functional}
{\cal F}_{ex}[\{\rho_i\}]=\int {\mathrm d}^3 r~ \Phi(\{n_\alpha({\bf r})\}),
\end{equation}
where $\Phi$ is a function of a set of weighted densities $\{n_\alpha\}$ which 
are defined by
\begin{equation} \label{weights}
n_\alpha({\bf r})=\sum_{i=1}^{\nu} \int {\mathrm d}^3 r' \rho_i({\bf r}')~
\omega_i^\alpha
({\bf r}-{\bf r}').
\end{equation}
The weight functions $\omega_i^\alpha$ in Eq.~(\ref{weights}) depend only on 
the geometrical features, the so-called fundamental measures,  of species $i$. 
Explicit expressions for the weight functions of hard-sphere mixtures and for 
$\Phi$ can be found in Ref.~\cite{Rosenfeld89} and in Ref.~\cite{Rosenfeld96}. 
The Rosenfeld functional has the following properties: i) the free-energy of 
the homogeneous mixture is identical to that from Percus-Yevick or 
scaled-particle theory and ii) the pair direct correlation functions of the 
homogeneous hard-sphere mixture, generated by functional differentiation of 
${\cal F}_{ex}$, are identical to those of Percus-Yevick theory. The index 
$\alpha$ labels 4 scalar plus 2 vector weights \cite{Rosenfeld89}. While the 
{\em original} functional given in Ref.\cite{Rosenfeld89} did  not account for
the freezing transition of pure hard spheres, more sophisticated extensions 
\cite{Rosenfeld96} do account for freezing; the weight functions remain the 
same but $\Phi$ is changed slightly. For the depletion potential problems 
under consideration the different versions give almost identical results for 
bulk packing fractions $\eta_s \lesssim 0.3$ \cite{comment}. At higher packing 
fractions the density profiles of the small spheres $\rho_s(z)$ close to a 
hard planar wall, or $\rho_s(r)$ close to a fixed particle, do display small 
deviations between the different versions of the Rosenfeld functional. 
Moreover, when calculating the depletion potential for size ratios of $s=0.1$ 
or smaller, these deviations are amplified and one observes slightly smaller 
amplitudes of oscillation for the more sophisticated versions of the theory. 
An example is given in Subsection~\ref{sec:oscil} (see, c.f., 
Fig.~\ref{fig:crocker}).

The one-body direct correlation function, defined within density functional 
theory by Eq.~(\ref{onebody}), can be written as
\begin{equation} \label{c1DFT}
c_b^{(1)}({\bf r};\{\mu_i\}) = -\sum \limits_\alpha \int {\mathrm d}^3 r' 
\left(\frac{\beta \partial \Phi(\{n_\alpha\})}
{\partial n_\alpha}\right)_{{\bf r}'} \omega_b^\alpha({\bf r}'-{\bf r})
\end{equation}
for the Rosenfeld functional. In the limit where all species have the same 
radius it is easy to check that the weighted densities $n_\alpha$ in 
Eq.~(\ref{weights}), and hence $\Phi$, reduce to the corresponding quantities 
for the pure fluid and, since the weight functions $\omega_b^\alpha$ in 
Eq.~(\ref{c1DFT}) reduce to the weight functions of the pure system, 
$c_b^{(1)}$ reduces to the one-body direct correlation function of the pure 
($s$) fluid. The depletion potential is then given by 
$\beta W(r) = - \ln(\rho_s(r)/\rho_s(\infty))$, which is the correct result
\cite{Goetzelmann98}.

Defining functions $\Psi^\alpha$ as
\begin{equation} \label{deriv}
\Psi^\alpha({\bf r}')\equiv \left(\frac{\beta \partial \Phi(\{n_\alpha\})}
{\partial n_\alpha}\right)_{{\bf r}'}-\left(\frac{\beta \partial
\Phi(\{n_\alpha\})} {\partial n_\alpha}\right)_{\infty},
\end{equation}
the grand potential difference in Eq.~(\ref{dep}) can be written as a sum of
convolutions of these functions with the weight functions of species $b$:
\begin{equation} \label{deppot}
\beta W_t({\bf r})=\sum_\alpha \int {\mathrm d}^3 r' \Psi^\alpha({\bf r}')~
\omega_b^\alpha({\bf r}'-{\bf r}).
\end{equation}

This expression is valid for arbitrary densities. The dilute limit of species 
$b$ can now be taken, within the Rosenfeld functional, by considering the 
weighted densities Eq.~(\ref{weights}) which in this case reduce to 
\begin{equation} \label{dilute_funk}
n_\alpha^{dilute}({\bf r})=\sum_{i\not=b} \int {\mathrm d}^3 r' 
\rho_i({\bf r}')~\omega_i^\alpha({\bf r}-{\bf r}'),
\end{equation}
where the set of density profiles $\{\rho_i({\bf r})\}$ that enter
Eq.~(\ref{dilute_funk}) is that of the $\nu -1$ component fluid, i.e., the one
obtained after taking the limit. It follows that the Helmholtz free energy in
Eq.~(\ref{functional}) and, consequently, the functions $\Psi^\alpha$ in 
Eq.~(\ref{deriv}) are those of a $\nu-1$ component mixture. Species $b$ enters
into the calculation of the depletion potential, i.e., the dilute limit of 
Eq.~(\ref{deppot}), only through its geometry, i.e.,  via the weight functions 
$\omega_b^\alpha$. 

This feature of the theory becomes especially important if the number of 
components $\nu$ is small. In the particular case of a binary mixture, 
$\nu=2$, the minimization of the functional in the dilute limit reduces to the 
minimization of the functional of a pure fluid and the weighted densities 
depend only on the density profile $\rho_s({\bf r})$ of the small spheres:
\begin{equation} \label{dilute_bin}
n_\alpha^{dilute}({\bf r})=\int {\mathrm d}^3 r' \rho_s({\bf r}')~
\omega_s^\alpha({\bf r}-{\bf r}').
\end{equation}
Although the direct approach of calculating the depletion potential via
evaluating grand potential differences by brute force requires only a 
functional describing the pure fluid, the above considerations demonstrate
that our present approach based on the one-body direct correlation function of
the big spheres in a sea of small ones requires a functional that describes 
the binary mixture. 

\section{Results from the DFT approach and assessment of their accuracy} 
\label{sec:test_acc}

In this section we examine the accuracy of some of the approximations inherent
in the present DFT approach by comparing our DFT results for the depletion 
potential with those of simulations and with the predictions of the general 
asymptotic theory given in Subsection~\ref{sec:asympt}.

\subsection{Consistency check}

We consider first the results of two separate routes to obtaining the dilute 
limit for the case of a binary hard-sphere mixture. In the first route both 
components of the mixture are treated on equal footing so that one calculates 
both $\rho_b({\bf r})$ and $\rho_s({\bf r})$ and obtains $W_t({\bf r})$ using
Eq.~(\ref{deppot}). By requiring the chemical potential of the big spheres 
$\mu_b$ to become more and more negative, the bulk density $\rho_b(\infty)$ of
this component approaches zero and the dilute limit is taken numerically. For
all the mixtures we investigated, a bulk packing fraction of the big spheres 
of $\eta_b=10^{-4}$ was sufficiently small to ensure that the density profile 
of the small spheres is indistinguishable from that of a pure fluid at the 
same $\eta_s$. Moreover the convergence of $\rho_s({\bf r})$ and 
$W_t({\bf r})$ to their limiting values is rather fast; an explicit example is
given in Fig.~1 of an earlier Letter~\cite{Goetzelmann99}. Using the second 
route, employing the weighted densities of Eq.~(\ref{dilute_bin}), the dilute 
limit is taken directly in the functional. In Fig.~\ref{fig:dilute} depletion 
potentials corresponding to both routes are shown for a big hard sphere near a
planar wall and a size ratio $s=R_s/R_b=0.1$. The bulk packing fraction of the
small spheres is $\eta_s=0.3$. We find excellent agreement between the two 
sets of results. The same level of agreement is found for a wide range of size
ratios $s$ and packing fractions $\eta_s$. From this we conclude that the 
limit can be taken directly in the functional, which makes the calculations 
significantly easier to perform, and all the results for the depletion 
potential we present subsequently will be based on this route.

\subsection{Comparison with simulation data}

The results presented in Fig.~\ref{fig:dilute} test the self-consistency of 
the two routes to the dilute limit within the given DFT approach. In order to 
test the accuracy of approximations introduced by employing the Rosenfeld 
functional the results of the present approach are compared with those of 
simulations. Fortunately some independent sets of simulation results for 
depletion potentials are available for both the sphere-sphere and the 
wall-sphere case. In Fig.~\ref{fig:biben} the depletion potentials between two
big spheres in a sea of small spheres, at a size ratio of $s=0.1$ and various 
packing fractions up to $\eta_s=0.6 \pi/6 \approx 0.314$, obtained from the 
molecular dynamics simulations of Ref.~\cite{Biben96} are compared with 
results of the present DFT approach. The agreement between the latter and the 
simulations is generally very good. At the higher packing fractions small 
deviations can be seen near contact and near the first minimum but the 
agreement is within the error bars of the simulations \cite{private} which
are not indicated here. We note 
that in Ref.~\cite{Biben96} the depletion {\em force} was the quantity 
measured in the simulations and the depletion potential was calculated by
integrating a {\em smoothed} force. For a higher packing fraction, 
$\eta_s=0.7 \pi/6 \approx 0.367$, the agreement between the depletion potential
obtained in the simulations of Ref.~\cite{Biben96} and our present result is 
poorer (not shown in Fig.~\ref{fig:biben}), but for this large value of 
$\eta_s$ the error bars of the simulations are probably bigger than for small 
values of $\eta_s$ \cite{private}.

In Ref.~\cite{Goetzelmann99} the depletion  potential for a single big 
hard sphere near a planar hard wall calculated within DFT was compared with 
the results of two independent sets of simulations for a size ratio $s=0.2$ 
and a packing fraction $\eta_s=0.3$. Very good agreement was found. In 
Fig.~\ref{fig:dickman} we present a comparison of our results with simulation 
results from Ref.~\cite{Dickman97} for size ratios $s=0.2$ (a) and $s=0.1$ (b),
for various packing fractions of the small spheres up to $\eta_s=0.3$. The 
original simulation results did not oscillate around $W=0$ which led us to
follow the procedure described in Ref.~\cite{Goetzelmann99} and to shift the 
data by a small constant amount in order to match the contact values with 
those of our DFT result. We note that in the simulations of 
Ref.~\cite{Dickman97} the depletion force was measured and the depletion 
potential was obtained by integrating the force. Since the data for the force
are available only for $h \leq h_{max}$, the integral depends on the cut-off
$h_{max}$.We surmise that this cut-off dependence is responsible for $W(h)$ 
not oscillating around zero. The agreement between our DFT results and those 
of the shifted simulation data is very good. The differences probably lie 
within the error bars of the simulations, for all packing fractions when 
$s=0.2$, and for $\eta_s=0.1$ and $\eta_s=0.2$ when $s=0.1$. However, for 
$\eta_s=0.3$ and $s=0.1$ clear deviations remain between our results and those 
of the simulations. In this case the shifted simulation data for the depletion
potential are close to the DFT results for $h<\sigma_s$ -- the height and 
position of the first maximum are the same -- but, in contrast to the DFT
results, the simulation data do not oscillate around zero. Clearly some 
alternative procedure for interpreting the simulation data is required.

\subsection{Density profiles}

In Fig.~\ref{fig:profiles} the number density profiles of a binary hard-sphere
mixture near a planar hard wall as obtained from DFT are shown for three size 
ratios. The packing fraction of the small spheres is $\eta_s=0.3$ and that of 
the big spheres is $\eta_b=10^{-4}$. The latter is sufficiently low that the 
density profiles of the small spheres, $\rho_s(z)$, shown in 
Fig.~\ref{fig:profiles}(a), are practically equal to that corresponding to
pure small spheres. Therefore, they are indistinguishable for all size ratios.
Because of the hard-body interaction between the small spheres and the wall 
the density profile $\rho_s(z)$ exhibits a discontinuous fall to zero at 
$z=\sigma_s/2$. The density profiles of the big spheres for size ratios $s=0.1$
(full line), 0.1333 (dotted line) and 0.2 (dashed line) are shown in 
Fig.~\ref{fig:profiles}(b). These density profiles do differ significantly for 
different size ratios. The hard-body interaction between the wall and the big 
spheres does not allow their centers to encroach closer than $z=\sigma_b/2$ 
and we find that the contact value is very different in all three cases (see 
caption to Fig.~\ref{fig:profiles}). We note that the wavelength of the 
oscillations in both $\rho_s(z)$ and $\rho_b(z)$ is approximately $\sigma_s$. 
In order to display the asymptotic behavior of these density profiles the 
logarithm of the difference between each density and its bulk value is shown 
in Fig.~\ref{fig:asympt1}. For $z/\sigma_s \gtrsim 2$ these plots conform very
closely to the asymptotic form given by Eq.~(\ref{lowdensprof2}). Straight
lines joining the maxima have a common slope and the distance between adjacent
maxima is the same in all cases. Only the amplitudes of the oscillations in 
$\rho_b(z)$ differ for different values of $s$. It follows that the decay 
length, $a_0^{-1}$, and the wavelength of the oscillation, $2 \pi/a_1$, are 
the same for {\em both} density profiles, i.e., for the big and the small 
spheres, and are independent of the size ratio. We have confirmed that
the same values for $a_0$ and $a_1$ are obtained from plots of the density 
profiles of the same binary mixture in the presence of a fixed big hard sphere,
i.e., our results are consistent with Eq.~(\ref{lowdensprof}). At high
packing fractions of the small spheres, e.g., $\eta_s=0.42$, we can easily
resolve up to 25 damped oscillations. At long range we find the calculated
density profiles to be in excellent agreement with the predictions of the 
theory of asymptotic decay. As the amplitude of the 24th oscillation is 
smaller than the amplitude of the first one by a factor of approximately 
$5\times 10^{-6}$ this attests further to the high numerical accuracy of our 
results. In addition we confirmed numerically that the modifications of the 
Rosenfeld functional which we employed lead to the same asymptotic behavior of
$W(h)$ as the original functional \cite{comment}.

\subsection{Asymptotic behavior} \label{sec:testasympt}

The asymptotic behavior of the depletion potential calculated within DFT is 
shown in Fig.~\ref{fig:asympt2}. For $z/\sigma_s \gtrsim 2$ our results 
conform very closely to Eq.~(\ref{asymptpot2}): although the amplitude of the 
oscillations depends on $s$, $W(z)$ is characterized by the same, common decay
length $a_0^{-1}$ and wavelength $2 \pi/a_1$ which describe the density 
profiles of the mixture. The results displayed in 
Figs.~\ref{fig:asympt1} and \ref{fig:asympt2} indicate that the asymptotic 
behavior of the density profiles and of the depletion potential set in at 
rather small distances from the wall. For wall-sphere surface separations of 
typically $z \sim 2 \sigma_s$, or even smaller, the asymptotic formulae are 
already remarkably accurate. This is in keeping with the results of earlier 
studies of the bulk pairwise correlation functions of hard-sphere mixtures 
\cite{Evans94}, where leading-order asymptotics were shown to be accurate down
to second-nearest neighbor separations. We shall make use of this observation 
in a later section in which we develop an explicit parameterized form for the 
depletion potential.

As a final examination of the validity of the asymptotic analysis we calculated
values of $a_0$ and $a_1$ from plots of (the logarithm of) the density profiles
of the small spheres $\rho_s(z)$ near the planar hard wall (see 
Fig.~\ref{fig:asympt1}) for a range of values for $\eta_s$ and various size 
ratios from $s=0.5$ to $s=0.1$. In accordance with the above statement the 
results for $a_0$ and $a_1$ do not depend on $s$ and are shown in 
Fig.~\ref{fig:a0a1} together with the values obtained 
using the Percus-Yevick result for $c_{ss}^{(2)}(r)$ in the pure fluid to 
solve Eq.~(\ref{condition}) for the poles $q=a_1+ i a_0$; we recall that the 
Rosenfeld hard-sphere functional generates the Percus-Yevick two-body direct 
correlation functions for a bulk mixture \cite{Rosenfeld89}. For small packing
fractions $\eta_s$ the oscillations are damped very rapidly, i.e., the decay 
length $a_0^{-1}$ is small, so that the numerical determination of the 
wavelength from a density profile is quite difficult. Nevertheless, the level 
of agreement between the two sets of results is very good, for all values of 
$\eta_s$ that were considered, confirming that the DFT results are consistent 
with the general predictions for the asymptotic behavior.

Our approach predicts depletion potentials for both the wall-sphere and the 
sphere-sphere case which are in very good agreement with simulations for 
distances close to contact and which are consistent with predictions of the 
general theory of the asymptotic decay of correlations in hard-sphere mixtures 
for distances away from contact. From our comparisons we conclude that our 
approach yields accurate results in the whole range of distances, for 
packing fractions up to (at least) $\eta_s=0.3$ and for size ratios down to
(at least) $s=0.1$. We emphasize that the full structure of the depletion potential, which is correctly described by the present approach, is not captured by 
the Asakura-Oosawa approximation or by a truncated virial expansion
\cite{Mao95a,Goetzelmann98}.

\subsection{Large asymmetries} \label{sec:asym}

So far it is not apparent how well our present approach will fare for extreme 
asymmetries, i.e., for $s \ll 1$. The Rosenfeld functional, which is the 
density functional we apply for all of the calculations of the depletion 
potentials, is designed to treat a multi-component hard-sphere mixture with 
arbitrary inhomogeneities. Whilst its accuracy in describing the density 
profiles for a pure fluid \cite{Rosenfeld89} and for binary mixtures 
\cite{Roth00} at moderate packing fractions and moderate size ratios has been 
confirmed by comparison with simulation results, a highly asymmetric binary 
mixture has not yet been studied systematically using this functional. Thus it 
is not known for which size ratios the results calculated with this functional 
are accurate. We recall that the Percus-Yevick approximation becomes 
increasingly less accurate for bulk properties as $s\to 0$, but here we are 
interested, in particular, with the reliability of our approach for 
determining depletion potentials. The latter are obtained from density 
profiles, having taken the dilute limit of one of the species [see 
Eq.~(\ref{dilute2})].

In this context it is instructive to consider the depletion potential between 
a hard sphere of radius $R_1$ and one of radius $R_2$ in a sea of small 
hard spheres of radius $R_s$ at a packing fraction $\eta_s$. This system is 
formally a mixture of three components in which two are dilute. The radius 
ratio $R_s/R_1$ is chosen such that on the basis of our previous results we 
know that the Rosenfeld functional can treat a mixture of species 1 and $s$ 
accurately. On the other hand, the radius $R_2$ is chosen to be much bigger 
than $R_s$ and $R_1$. The depletion potential can be calculated in two 
different ways. In the first route, sphere $2$ (with large radius $R_2$) is 
fixed and enters into the calculation as an external potential for sphere 1 
and species $s$. The density profile of the small spheres in the presence of
this external potential can be calculated, and from it the depletion potential,
using the theoretical approach described in Sec.~\ref{sec:theory}. Thus, in 
this calculation the Rosenfeld functional treats a mixture with a moderate 
size ratio $R_s/R_1$ exposed to an external potential. Therefore we expect 
these results to be very accurate. In the second route, the roles of spheres 1
and 2 are exchanged. The sphere of medium radius $R_1$ is fixed and acts as an
external potential for the very large sphere 2 and the small species $s$. Now 
the Rosenfeld functional must treat a very asymmetric mixture. Of course, in 
an {\em exact} treatment of this problem it does not matter which sphere is 
fixed first as the depletion potential is simply the difference in the grand 
potential between a configuration in which spheres 1 and 2 are fixed and 
positioned close
to each other and one in which both spheres are at infinite separation. The 
result of an exact treatment cannot depend upon which sphere is regarded as an
external potential. However, it is not immediately obvious that the underlying
symmetry is respected in an approximate DFT treatment. At first sight one way 
of calculation might appear to be less demanding on the theory than the other.

In Fig.~\ref{fig:test} we show the depletion potential between the big spheres 
when sphere $2$ is fixed (solid line) and then with sphere $1$ fixed (symbols),
for a packing fraction of the small spheres $\eta_s=0.3$, and $R_1=5 R_s$. Two
different values of $R_2$ are considered, namely $R_s/R_2=0.02$ (a) and 
$R_s/R_1=0.01$ (b). We find excellent agreement between the results of the two 
routes for both (a) and (b). Only very small differences between the curves 
can be ascertained and these occur for separations close to contact where 
numerics are most difficult. Note that the results in 
Figs.~\ref{fig:test}(a) and \ref{fig:test}(b) lie very close to each other.
This can be understood easily when $R_2$ is the fixed sphere. For cases (a) 
and (b), $R_2 \gg R_1=5 R_s$ and one is effectively in the planar-wall limit 
so that both sets of results lie close to those in Fig.~\ref{fig:dickman}(a),
with $\eta_s=0.3$. From the results shown in Fig.~\ref{fig:test} and further 
comparisons for other values of $\eta_s$ it is evident that the Rosenfeld 
functional does maintain the required symmetry between 1 and 2. It is 
important to understand this. In the low density limit, i.e., if the ternary 
mixture is considered with the density of the small spheres also approaching 
zero, the depletion potential can be expressed in terms of Mayer $f$ functions
and the equivalence of the two routes can be verified directly. Starting from 
the functional given in Eq.~(\ref{lowdensfunc}) we follow the derivation of 
Eq.~(\ref{lowdensity}) and obtain 
\begin{equation}
\beta W_{21}(r) = -\rho_s(\infty) \int {\mathrm d}^3 r' f_{2s}({\bf r}') 
f_{s1}({\bf r}-{\bf r}'),
\end{equation}
for the depletion potential with sphere 2 fixed and
\begin{equation}
\beta W_{12}(r) = -\rho_s(\infty) \int {\mathrm d}^3 r' f_{1s}({\bf r}') 
f_{s2}({\bf r}-{\bf r}')
\end{equation}
for that with sphere 1 fixed. Here $f_{1s}$ and $f_{2s}$ are the Mayer $f$ 
functions between a small sphere and sphere 1 and 2, respectively, and it is 
evident that $W_{21}(r) \equiv W_{12}(r)$. The Rosenfeld functional will 
reproduce this result for packing fractions $\eta_s \to 0$, since it reduces 
to Eq.~(\ref{lowdensfunc}) in this limit. For arbitrary values of $\eta_s$ it 
is necessary to reconsider the genesis of the functional and recognize that 
although the hard-sphere pairwise potentials $\Phi_{ij}(r)$ between species 
$i$ and $j$ do not enter explicitly, the functional does respect the 
equivalence of $\Phi_{ij}(r)$ and $\Phi_{ji}(r)$; the Mayer functions and the 
weight functions which were used in constructing the functional are symmetric
w.r.t. $i$ and $j$. It is straightforward to show that the equivalence of 
$W_{21}(r)$ and $W_{12}(r)$ is guaranteed provided the functional respects this
symmetry. Thus, the two sets of results shown in Fig.~\ref{fig:test}(a) 
{\em should} agree with each other, as should those shown in 
Fig.~\ref{fig:test}(b). That there are small discrepancies reflects only
numerical inaccuracies rather than any fundamental shortcoming of the DFT 
approach. It is pleasing that what appear to be two distinct ways of 
calculating the depletion potential yield the same results, even for high
degrees of asymmetry. Whether other functionals, not based on fundamental
measure theory, will respect the symmetry requirements remains to be 
ascertained. Although the present calculations should be regarded as
a further test of the internal consistency of our approach rather than a formal
demonstration that it is accurate for extreme asymmetries, the results, when
coupled with the excellent agreement between theory and simulations for
$s=0.1$, do suggest that the approach should remain accurate for smaller 
size-ratios.

\section{Derjaguin approximation} \label{sec:derjag}

In the well-known Derjaguin approximation \cite{Derjaguin34} the force between 
two large convex bodies is expressed in terms of the interaction energy of two 
parallel plates. This approximate mapping is valid in the limit where the 
minimal separation of surfaces $h$ is much smaller than the radii of curvature 
and was developed assuming the force between the surfaces can be calculated by
integration over all interactions between pairs of points of the two bodies.
Recently the Derjaguin approximation was implemented for the depletion force 
between two big hard spheres in a sea of small hard spheres, employing a 
truncated virial expansion to calculate the excess pressure of the small 
spheres between planar hard walls, and results were compared with simulation 
data for a size ratio $s=0.1$ \cite{Mao95a}. There is, however, an important 
conceptual difference from earlier applications of the Derjaguin approximation
as depletion effects are global effects arising from packing of the small 
spheres and it is not obvious that the original derivation remains applicable 
or what the regime of validity of the approximation should be. Some of its 
limitations were discussed in Ref.~\cite{Goetzelmann98} where it was argued 
that the Derjaguin approximation should not be reliable for $s=0.1$ if the 
packing fraction $\eta_s \gtrsim 0.3$. Here we examine some of the key 
predictions of the Derjaguin approximation by making comparison with results 
of our DFT approach. From the arguments of Subsec.~\ref{sec:asym} it is safe 
to assume that the present DFT approach remains reliable for rather large size 
ratios where the Derjaguin approximation might be expected to be valid.

There is an elegant scaling relation connecting the depletion force,
$F(h)\equiv - \partial W(h)/\partial h$, between two big spheres, $F_{bb}(h)$, 
in a sea of small spheres with that between a single big sphere and a planar 
hard wall, $F_{wb}(h)$. In the limit of infinite asymmetry, $s\to 0$, the 
forces are equal except for a factor of $2$, i.e., 
\begin{equation} \label{scaling}
2 F_{bb}(h) = F_{wb}(h),
\end{equation}
with $h$ the minimal separation of the surfaces of the two big objects. This 
scaling relation follows directly from the Derjaguin approximation and if 
it is found to be obeyed it is sometimes inferred \cite{Attard89,Attard90} 
that the Derjaguin approximation itself is valid. However, it was shown 
\cite{Goetzelmann98} that this scaling relation follows from geometrical
considerations without introducing the explicit Derjaguin approximation. This 
can be illustrated by comparing the explicit Asakura-Oosawa depletion 
potentials [see Eqs.~(\ref{asabb}) and (\ref{asawb})]. In the limit 
$R_b \gg R_s$ both formulae reduce to 
$-\frac{\varepsilon}{2} \rho_s(\infty) \pi R_b (2 R_s -h)^2$, for  $h<2 R_s$, 
where $\varepsilon=1$ corresponds to the sphere-sphere and $\varepsilon=2$ to
the wall-sphere case. Thus, achieving the correct scaling property in 
Eq.~(\ref{scaling}) does not prove that the Derjaguin approximation,
\begin{equation} \label{derj}
F^{Derj}(h) = - \varepsilon \pi (R_b + R_s) \int \limits_h^\infty {\mathrm d}L~
f_s(L)
\end{equation}
is accurate. Here $f_s(L)$ is the solvation force, or the excess pressure, for
the small-sphere fluid confined between two planar parallel hard walls 
separated by a distance $L$ \cite{Goetzelmann98}.

From our DFT calculations we find that the scaling relation is already 
well-obeyed at moderate size ratios. In Fig.~\ref{fig:scaling} the scaled 
depletion force $\beta f^*_{bb}(h)=2 \beta F_{bb}(h) R_s^2/(R_b+R_s)$
between two big hard spheres (solid line) and that between a single big hard 
sphere and a planar hard wall ($\square$),
$\beta f^*_{wb}(h) = \beta F_{wb}(h) R_s^2/(R_b+R_s)$, in a sea of small 
hard spheres at a packing fraction of $\eta_s=0.3$ is shown for size ratios 
$s=0.1$ and 0.02. While small deviations from the scaling relation in 
Eq.~(\ref{scaling}) are visible close to contact for $s=0.1$, these deviations 
have almost disappeared for $s=0.05$ (not shown in the figure) and near 
perfect agreement is found for $s=0.02$. Note also that the scaled depletion 
forces corresponding to the different values of $s$ lie close to each other.

An explicit result of the Derjaguin approximation [Eq.~(\ref{derj})] is that 
the depletion force between two big spheres or between a big sphere and a 
planar wall can be written as \cite{Goetzelmann98}
\begin{equation} \label{derjforce}
F^{Derj}(h) = \varepsilon \pi (R_b+R_s) (p(\eta_s) (h - 2 R_s) - 
\gamma(\eta_s)),~~h<2 R_s,
\end{equation}
where $p(\eta_s)$ is the bulk pressure of the small spheres and
$\gamma(\eta_s)$ is twice the surface tension of the small-sphere fluid at a 
planar hard wall. The geometrical factor $\varepsilon$ is the same as in 
Eq.~(\ref{derj}). Thus, for a given size ratio the slope of the depletion 
force predicted by the Derjaguin approximation is constant for $h<2 R_s$ and 
depends only on the equation of state of the small spheres $p(\eta_s)$. For 
the particular case of hard spheres we obtain:
\begin{equation} \label{slope}
\frac{{\mathrm d} \beta F^{Derj}(h)}{{\mathrm d} h} 
= \varepsilon (R_b+R_s) \frac{3 \eta_s}{4 R_s^3} \frac{1+\eta_s+\eta_s^2-
\eta_s^3}{(1-\eta_s)^3},~~h<2 R_s,
\end{equation}
where the quasi-exact Carnahan-Starling equation of state\cite{Carnahan69} was 
used. 

However, the depletion forces calculated within the present approach show a
qualitatively different behavior from that predicted by Eq.~(\ref{slope}). It 
was found that even for small size ratios ($s \leq 0.05$), only in the limit 
$\eta_s\to 0$, in which the Asakura-Oosawa approximation becomes exact, there
is agreement between the Derjaguin approximation and the results of our 
approach. The depletion force calculated at a packing fraction of $\eta_s=0.3$ 
does not have constant slope for $h<2 R_s$ (see Fig.~\ref{fig:scaling}). This 
is in clear contradiction to Eq.~(\ref{slope}). Simulation results 
\cite{Dickman97} for the depletion force also exhibit non-constant slopes for 
$h<2 R_s$.

Another prediction of the Derjaguin approximation in Eq.~(\ref{derjforce}) is 
that the contact value $W(0)$ of the depletion potential can be expressed 
simply in terms of the equation of state $p(\eta_s)$ and the surface tension 
$\gamma(\eta_s)$. Using the Carnahan-Starling result for $p(\eta_s)$ 
\cite{Carnahan69} and the scaled particle result for $\gamma(\eta_s)$ 
\cite{Reiss60}  we obtain \cite{Goetzelmann98}:
\begin{equation} \label{contactvalue}
\beta W^{Derj}(0) = -\frac{\varepsilon (R_b+R_s) 3 \eta_s}{2 R_s}
\frac{1-2 \eta_s -2\eta_s^2-\eta_s^3}{(1-\eta_s)^3}
\end{equation}
which becomes positive at high packing fractions of the small spheres 
\cite{Goetzelmann98}. Provided $\eta_s<0.2$ the contact values from
Eq.~(\ref{contactvalue}) are in reasonable agreement with the results of our 
DFT approach for small values of $s$. However, the contact values obtained 
from Eq.~(\ref{contactvalue}) change sign at $\eta_s \approx 0.3532$, which is
in complete contradiction to the results of the present approach 
where we find negative contact values for all packing fractions and all size 
ratios $s$ under consideration \cite{Corti98}.

In Ref.~\cite{Goetzelmann98} it was shown that a third-order virial expansion 
(in powers of $\eta_s$) for the depletion potential calculated within the 
Derjaguin approximation does not yield positive contact values. However, 
expansion to fourth or fifth-order shows a qualitatively different behavior 
from third order and already indicates the onset of positive $W(0)$. Thus, in 
keeping with Ref.~\cite{Goetzelmann98} we conclude that the Derjaguin 
approximation is not very useful for the calculation of depletion forces. The 
good level of agreement, observed for $h<2 R_s$, between the results of the 
third-order virial expansion \cite{Mao95a} and those of simulation 
\cite{Biben96} for $s=0.1$ should be regarded fortuitous.

\section{Applications} \label{sec:appl}

\subsection{A parametrized form for the depletion potential of hard spheres}
\label{sec:param}

As mentioned in Subsec.~\ref{sec:general}, recent studies of correlation
functions and phase equilibria of highly asymmetric binary mixtures have shown
that it is very advantageous to map such mixtures onto effective one-component
fluids \cite{Dijkstra98,Dijkstra99}. The effective pairwise potential between 
the big particles is then the bare pair potential between two big particles
plus the depletion potential [see Eq.~(\ref{phitot})]. Thus, in calculating 
the phase behavior of binary hard-sphere mixtures it is necessary to adopt 
a specific form for the depletion potential between two big hard spheres. 
Previous simulation studies \cite{Dijkstra98} of binary mixtures have employed
the simplified third-order virial expansion formula given by G{\"o}tzelmann et 
al. \cite{Goetzelmann98} and the same potential has been used in a perturbation
theory treatment of the phase behavior \cite{Dijkstra99}. Although this 
formula is convenient for global investigations of phase behavior, as the 
depletion potential is given explicitly as a function of $\eta_s$, clearly it 
would be valuable to have a simple, parametrized form for the depletion 
potential that (i) is better founded than the formula provided by
G{\"o}tzelmann et al. and (ii) captures the correct intermediate and 
long-range oscillatory structure as well as the important short-range features.
Note that in Refs.~\cite{Dijkstra98} and \cite{Dijkstra99} the effective pair 
potential was set equal to zero for separations $h>2 R_s$.

We have used depletion potentials calculated within the present DFT approach, 
for a single big sphere near a planar hard wall and for two big spheres, to 
develop a suitable parameterization scheme. Although this parameterization is 
fairly simple it yields rather accurate fits. The depletion potential close to 
contact is fitted by a polynomial and is continued by the known 
asymptotic behavior. In the following the variable $x$ measures the minimal 
distance from contact in units of the small sphere diameter $\sigma_s$, i.e., 
$x\equiv h/\sigma_s$. These parametrized depletion potentials $\bar W$ are also
scaled: the actual potentials $W$ are recovered by multiplying by a factor of 
$\varepsilon (R_b+R_s)/(2 R_s)$ with $\varepsilon=2$ for the wall-sphere and 
$\varepsilon=1$ for the sphere-sphere potential:
\begin{equation} \label{scaldep}
W = \frac{\varepsilon (R_b+R_s)}{2 R_s} \bar W.
\end{equation}

Between contact at $x=0$ and the location $x_0$ of the first maximum the scaled
depletion potential is fitted by a cubic polynomial:
\begin{equation} \label{fit}
\beta \bar W(x,\eta_s) = a(\eta_s) + b(\eta_s)~ x + c(\eta_s)~ x^2 + 
d(\eta_s)~ x^3,~~x<x_0,
\end{equation}
where the coefficients $a$, $b$, $c$ and $d$ are functions of the packing 
fraction of the small spheres $\eta_s$. More details of this polynomial and the
determination of the coefficients are presented in the appendix.

In order to obtain the depletion potential for $x>x_0$ we assume that the 
asymptotic decay already sets in at the point $x_0$. This assumption is 
supported by the results presented in Fig.~\ref{fig:asympt2}. Thus, for 
$x>x_0$ we adopt the form  [c.f. Eq.~(\ref{lowdensprof2})]
\begin{equation} \label{asymptfitw}
\beta \bar W_{asympt}^w(x,\eta_s) = A_w(\eta_s) \exp(-a_0(\eta_s) \sigma_s x) 
\cos(a_1(\eta_s) \sigma_sx - \Theta_w(\eta_s)),~~x>x_0,
\end{equation}
for the scaled depletion potential between a wall and a sphere and [c.f.
Eq.~(\ref{lowdensprof})]
\begin{equation} \label{asymptfitp}
\beta \bar W_{asympt}^p(x,\eta_s) = \frac{A_p(\eta_s)}{s^{-1}+x}~ 
\exp(-a_0(\eta_s) \sigma_s x) \cos(a_1(\eta_s) \sigma_s x - \Theta_p(\eta_s)),
~~x>x_0,
\end{equation}
for the potential between two spheres. The denominator in 
Eq.~(\ref{asymptfitp}) measures the separation $\sigma_b+h$ between the 
centers of the spheres in units of $\sigma_s$. Both forms contain the 
functions $a_0(\eta_s)$ and $a_1(\eta_s)$, which can be calculated from the 
Percus-Yevick bulk pair direct correlation function $c^{(2)}_{ss}(r)$ (see 
Subsec. \ref{sec:testasympt} and Fig.~\ref{fig:a0a1}). The amplitudes 
$A_j(\eta_s)$ and phases $\Theta_j(\eta_s)$, $j=p,w$, are chosen so that the 
depletion potential and its first derivative are continuous at $x_0$. 
$A_p(\eta_s)$ and $\Theta_p(\eta_s)$ are weakly dependent on the size ratio 
$s$.

With this prescription the scaled depletion potential is completely determined.
For a given packing fraction $\eta_s$ the coefficients $a$, $b$, $c$ and $d$ 
are given by Eq.~(\ref{coeff}) and the position of the first maximum can be 
calculated from Eq.~(\ref{x0}). Using those values as input, the amplitude $A$ 
and the phase $\Theta$ of the asymptotic decay are readily obtained from either
Eqs.~(\ref{amplitude}) and (\ref{Theta}) or Eqs.~(\ref{amplitudep}) and 
(\ref{Thetap}). Thus in this parametrization the scaled depletion potential has
the form
\begin{equation} \label{depfit}
\beta \bar W(x,\eta_s) = \left\{
\begin{array}{rl}
a + b~ x + c~ x^2 + d~ x^3, & x\leq x_0\\
\beta \bar W_{asympt}^{p,w}(x,\eta_s), & x> x_0.
\end{array}
\right.
\end{equation}

In Fig.~\ref{fig:fit}(a) fits (lines) of the form given by Eq.~(\ref{depfit}) 
are compared with the scaled depletion potentials between a big hard sphere and
a hard wall calculated within DFT (symbols) for a size ratio $s=0.1$. Although
the fit is relatively simple, its accuracy is high. The position of the first 
maximum, which depends sensitively on the packing fraction $\eta_s$, is 
reproduced very accurately. The value $\beta \bar W_0 = \beta W(x_0)$ of the 
potential at the first maximum is also given quite accurately, and only for 
$\eta_s=0.3$ are small deviations of the fit from the full DFT results visible.
Clearly the full structure of the depletion potential is reproduced well by 
this parametrization. In order to demonstrate the wide range of applicability 
of this parametrization in Fig.~\ref{fig:fit}(b) we show a comparison of the 
parametrized scaled depletion potential ($\square$) for a packing fraction 
$\eta_s=0.3$ with scaled DFT results (lines) for the depletion potential 
between two spheres and size ratios $s=0.2$ and $s=0.05$. Although for these 
size ratios the scaling relation Eq.~(\ref{scaling}) is not satisfied 
particularly accurately, the agreement between our parametrization and the DFT 
results is rather good. This gives us confidence that we have developed a 
satisfactory parametrized form for the depletion potential which properly 
incorporates all essential features.

\subsection{Oscillatory depletion potential at high packing fractions of the 
small spheres} \label{sec:oscil}

In a recent experiment by Crocker et al. \cite{Crocker99} the equilibrium 
probability distribution $p(r)$ for two (big) PMMA (polymethylmethacrylate) 
spheres of diameter $\sigma_b=1.1 \mu$m immersed in a sea of (small) 
polystyrene spheres of diameter $\sigma_s=83$ nm was measured using 
line-scanned optical tweezers and digital videomicroscopy at various packing 
fractions in the range between $\eta_s=0.04$ and $\eta_s=0.42$. The solvent 
contains added salt and surfactant to prevent colloidal aggregation and the 
`bare' interactions between the colloidal particles are expected to be 
screened Coulombic repulsion with a screening length of about 3 nm 
\cite{Crocker99}. Since the latter is small compared with the colloid diameters
the bare interactions can be regarded, to good approximation, as 
hard-sphere-like. The depletion potentials $\beta W(r)=-\ln(p(r)/p(\infty))$ 
(see Subsec.~\ref{sec:theory} A) obtained from these experiments are shown in 
Fig.~1 of Ref.~\cite{Crocker99}. At low packing fractions, $\eta_s=0.04$ and 
0.07, rather good agreement with the results of the Asakura-Oosawa 
approximation was found, after taking into account the effects of limited
spatial resolution of the optical instruments. For $\eta_s=0.15$ and 0.21 the 
measured depletion potential displayed a pronounced repulsive barrier. For 
higher packing fractions, i.e., $\eta_s=0.26$, 0.34 and 0.42 damped 
oscillations were observed, these being particularly pronounced for the two 
highest packing fractions for which three maxima are clearly visible. 
Reference~\cite{Crocker99} appears to be the first report of an experimental 
observation of an oscillatory depletion potential and, indeed, of a repulsive 
contribution arising from purely entropic or packing effects \cite{comment2}.

Motivated by these experiments we consider a binary hard-sphere mixture in the 
dilute limit with a size ratio $s=0.0755$, as in the experiment. (We do not 
attempt to include the increase of the effective radius of the spheres arising
from screened Coulomb repulsion and, in keeping with the authors of 
Ref.~\cite{Crocker99}, we do not include any dispersion forces.) As previously,
the depletion potential between two big spheres is calculated using 
Eq.~(\ref{deppot}) in the dilute limit. The functions $\Psi^\alpha$ are 
functionals of $\rho_s(r)$, the density profiles of the small spheres close to
a big sphere fixed at the origin, which depend only on the radial distance $r$.
The results are shown in Fig.~\ref{fig:crocker} for the same values of $\eta_s$
as in the experiments. It is encouraging to find that the theoretical and 
experimental results have many common features. As expected, the calculated 
oscillations become much more pronounced as $\eta_s$ increases. The wavelength
decreases slowly and the decay length of the envelope increases rapidly with 
$\eta_s$ -- as predicted by the theory of asymptotic decay (see 
Fig.~\ref{fig:a0a1}). The experimental data are consistent with both 
observations. Moreover the wavelength of the oscillations for $\eta_s=0.34$ is
close to $\sigma_s=$ 83 nm in theory and experiment. For $\eta_s=0.42$ both 
theory and experiment yield a slightly smaller wavelength. The amplitude of 
the calculated oscillations is larger than in the experiment. However, we 
emphasize that we made no attempt to take into account effects of instrumental 
resolution or the polydispersity of the small polystyrene particles. Nor have 
we attempted to include the effects of the softness of the inter-particle 
potentials and any non-additivity of the effective diameters; both are likely 
to lead to a reduction in the amplitude of the oscillations. The qualitative 
agreement between the experimental results and those of our calculations 
persuades us that the hard-sphere model is an appropriate starting point for 
describing the colloidal system and that the observed oscillations do reflect 
the packing of the small spheres -- as inferred in Ref.~\cite{Crocker99}.

Significant deviations between our results and the experimental ones do occur, 
at large $\eta_s$, for separations near contact or near the first maximum in 
the depletion potential. Our results imply that the height of the first maximum
and the magnitude of the contact value $|W(0)|$ are larger than the 
experimental ones by about a factor of two for $\eta_s=0.34$. Although the 
source of these differences may well reside in the experimental situation it 
is important to check that the particular DFT which we employ is performing 
reliably at these high values of $\eta_s$. It is precisely this regime of high
density and very strong confinement of the small spheres where differences 
between the various DFT theories, i.e., the improvements on the original 
Rosenfeld version, might reveal themselves. These circumstances are reminiscent
of those investigated by Gonz{\'a}lez et al. \cite{Gonzalez98} in their DFT 
studies of hard spheres in small spherical cavities. Those authors were able 
to ascertain that the improved theories fared better than the original version
under conditions of extreme confinement. 

To this end we repeated our calculations of the depletion potential with the
improved versions of the Rosenfeld functional that can account for the 
freezing transition \cite{comment}. At packing fractions $\eta_s \lesssim 0.3$
we obtained, as stated earlier, results almost identical to those of the 
original functional. At higher packing fractions, however, we find that the 
amplitude of the oscillations is slightly smaller than those obtained from the
original functional. This is illustrated in Fig.~\ref{fig:crocker} for
$\eta_s=0.42$, using the interpolation form of the functional \cite{comment} 
(dotted line). The antisymmetrized version of the functional, with $q=3$
\cite{comment}, yields a depletion potential very close to that of the
interpolation form. In view of the smallness of these deviations the 
discrepancies between the experimental findings and the theoretical results
at high $\eta_s$ cannot be blamed on the performance of the DFT but most
probably reside in differences between the actual experimental sample and the
model of hard spheres.

\section{Summary and Discussion} \label{sec:conc}

In this paper we have developed a versatile theory for determining the
depletion potential in general fluid mixtures. Our approach requires only the
knowledge of the equilibrium density profile $\rho_s({\bf r})$ of the small 
particles {\em before} the big (test) particle is inserted, i.e., 
$\rho_s({\bf r})$ has the symmetry of the external potential. If the latter is
exerted by a fixed particle or by a planar wall then in these cases 
$\rho_s({\bf r})$ simplifies to functions $\rho_s(r)$ or $\rho_s(z)$ of one
variable. Since a one-dimensional profile can be calculated very accurately, 
the resulting depletion potentials can be obtained without the numerical 
complications and limitations that are inherent in brute-force DFT 
\cite{Frink00}. The latter requires the calculation of the local density of the
small particles around the big particles in the presence of the external 
potential \cite{Melchionna00} or the calculation of the total free energy as a 
function of the separation of the big particles \cite{Gruenberg99}; both 
calculations require considerable numerical effort due to the reduced symmetry 
of the density distributions. We have employed our approach in a comprehensive 
study of the depletion potential for hard-sphere systems, using Rosenfeld's 
fundamental measure functional. The main conclusions which emerge from our 
study are as follows:
\begin{enumerate}
\item The depletion potential can be obtained by considering a liquid mixture 
in the limit of vanishing concentration of one of the species. Two different 
ways to implement this limit lead to the same result (Fig.~\ref{fig:dilute}).
\item Detailed comparison of our results with those of simulations, for both
sphere-sphere and (planar) wall-sphere depletion potentials (see 
Figs.\ref{fig:biben} and \ref{fig:dickman}), demonstrate that the theory is 
very accurate for size ratios $s=R_s/R_b$ as small as 0.1 and for packing 
fractions $\eta_s$ as large as 0.3. These are the most extreme cases for which
reliable simulation data are presently available. The theory describes 
accurately the short-ranged depletion attraction, the first repulsive barrier 
and the subsequent oscillations in the depletion potential.
\item By performing consistency checks we argue that at least up to moderate
packing fractions the predictions of the Rosenfeld DFT for depletion should be
quantitatively reliable even for large asymmetries between the sizes of the
solvent and the solute particles (Fig.~\ref{fig:test}). 
Subsection~\ref{sec:asym} provides a theoretical understanding of this feature
of our DFT approach.
\item Extensions of the Rosenfeld functional \cite{comment} yield very
similar results (see Fig.~\ref{fig:crocker}) for the cases we have studied.
It would be of considerable interest to test the performance of the
proposed functionals against simulation data for smaller size ratios and for 
higher values of $\eta_s$, for which more extreme packing constraints 
{\em might} discriminate between the various functionals.
\item Our DFT approach incorporates the correct, exponentially damped,
oscillatory asymptotic ($h\to\infty$) decay of the depletion potential $W(h)$. 
This is inherent in the construction of the theory, is preserved by the
approximate Rosenfeld functional and is exhibited explicitly by the numerical
results (Fig.~\ref{fig:asympt2}). The decay length $a_0^{-1}$ of the
oscillations increases and the wavelength $2 \pi/a_1$ decreases with increasing
$\eta_s$ (Fig.~\ref{fig:a0a1}) but these quantities are independent of the 
size ratio $s$. The same values for $a_0$ and $a_1$ characterize the 
oscillatory decay towards the bulk values of the number density profiles of
hard-sphere mixtures near a hard wall when the packing fraction of the big
spheres is very small (Figs.~\ref{fig:profiles} and \ref{fig:asympt1}).
\item We have developed simple parametrization schemes for the depletion
potential between big hard sphere and a planar wall and that between two
big hard spheres which provide accurate fits to our DFT results (see
Fig.~\ref{fig:fit}). The fitting procedure makes use of the fact that 
leading asymptotic behavior of $W(h \to\infty)$ provides an accurate account of
the oscillatory structure of the depletion potential at intermediate 
separations as well as at longest range. Such parametrizations are designed to 
provide a more accurate alternative to the third-order virial expansion 
formula given by G{\"o}tzelmann et al. \cite{Goetzelmann98}. Since these
parametrization can be easily implemented we recommend that they should be 
employed in subsequent studies of the phase behavior of highly-asymmetric 
binary hard-sphere mixtures of the type reported in Refs.~\cite{Dijkstra98}
and \cite{Dijkstra99}.
\item In Sec.~\ref{sec:derjag} we investigated the regime of validity of the 
Derjaguin approximation [Eq.~(\ref{derj})] for the depletion potential and
showed that this fails, for all but the smallest packing fractions $\eta_s$,
for which the depletion potential reduces to the Asakura-Oosawa result. 
However, the scaling relation Eq.~(\ref{scaling}) connecting the depletion 
force between two big spheres to that between a big sphere and a planar wall
-- which is predicted by the Derjaguin approximation but which also follows 
from geometrical considerations -- does remain valid even at moderate size 
ratios (Fig.~\ref{fig:scaling}).
\end{enumerate}
We conclude with several remarks concerning the accuracy and usefulness of our
approach. One might be surprised that a DFT which corresponds to the
Percus-Yevick theory for the bulk mixture (the Rosenfeld functional yields the 
same bulk free-energy density and bulk pair direct correlation 
function) performs so well for small size ratios, for which it is known that 
Percus-Yevick theory becomes inaccurate. For example, it fails to predict the 
fluid-fluid spinodals for hard-sphere mixtures. However, our present approach 
involves only the calculation of a one-body direct correlation function 
$c_b^{(1)}({\bf r};\{\mu_i\})$ and, therefore, the determination of one-body 
density profiles. The minimization of approximate functionals can yield rather
accurate one-body profiles in spite of the limitations of the underlying
approximations; e.g., this is the reason why the test particle route to the 
bulk radial distribution function $g(r)$ is very successful within DFT 
\cite{Rosenfeld89,Evans92}. Furthermore, in determining the depletion 
potential $W({\bf r})$ we require only solutions of the Euler-Lagrange 
equation for $\rho_s({\bf r})$ in the limit where $\rho_b\to 0$, i.e., in the 
absence of the big particles. The DFT is likely to be more accurate in this 
limiting regime than for a mixture concentrated in all species. We emphasize 
that taking the dilute limit of the big particles numerically, i.e., working 
at non-zero but very small values of $\eta_b$, involves more computation than 
taking the limit directly in the functional. Moreover, caution should be 
exercised in hard-sphere mixtures with extreme size ratios, $s\leq 0.1$, at 
high packing fractions $\eta_s$ of the small spheres since the fluid-solid 
phase boundary already occurs at very low packing fractions $\eta_b$ of the 
big spheres \cite{Dijkstra98}. The fluid-solid coexistence region is avoided 
if the dilute limit is taken directly \cite{comment3}.

Our procedure for calculating $W_t({\bf r})$ at arbitrary concentrations of the
big particles might prove useful for interpreting (future) measurements of
the effective interaction potential when the mixture is not in the dilute 
limit. Figure~1 of Ref.~\cite{Goetzelmann99} illustrates how the wall-sphere 
potential $W_t(z)$ varies with the big sphere packing fraction $\eta_b$ for a 
mixture with size ratio $s=0.2$ and $\eta_s=0.2$. For $\eta_b=0.025$, $W_t(z)$ 
already differs by a few percent from its dilute limit $W(z)$.

It is possible to calculate the depletion potential by using as input density 
profiles obtained by other means. In particular one might take simulation data
for $\rho_s({\bf r})$, computed in the absence of the big test particle, and 
insert these into Eq.~({\ref{dilute_bin}) to determine the weighted densities.
Although such a procedure does not offer the appeal of a self-consistent 
approach in which both the equilibrium density profiles and the depletion 
potential are calculated within the same framework, in practice this could be
a profitable route for complex geometries where a direct simulation of the 
depletion potential or force is very difficult.

Finally we mention that the techniques we have developed here are not 
restricted to additive, binary hard-sphere mixtures. Our general approach to 
the calculation of depletion potentials can be applied to hard-sphere mixtures
with non-additive diameters, to ternary mixtures and to systems where the
interparticle potentials are soft.

\acknowledgements

We thank T. Biben and R. Dickman for providing us with their simulation data.
We benefited from conversations with J.M. Brader, M. Dijkstra, H.N.K 
Lekkerkerker, R. van Roij, and correspondence with J.-P. Hansen, P.B. Warren,
and A.G. Yodh. This research was supported in part by the EPSRC under 
GR/L89013.

\appendix

\section{Parameterizing the depletion potential}

In the range between contact, $x=h/\sigma_s=0$, and the position $x_0$ of the
first maximum the scaled depletion potential $\bar W$ is parametrized by a 
cubic polynomial [Eq.~(\ref{fit})], which is the simplest polynomial fit that 
remains accurate close to contact. Since $\beta \bar W(x=0,\eta_s)=a(\eta_s)$, 
the first coefficient is the contact value of the depletion potential. The 
position $x_0$ and the height $W_0$ of the first maximum can be obtained 
easily by differentiating Eq.~(\ref{fit}). The cubic polynomial has two 
extrema, with the maximum located at
\begin{equation} \label{x0}
x_0(\eta_s) = - \frac{c + \sqrt{c^2-3 b d}}{3 d}
\end{equation}
and a maximal value of
\begin{equation}
\beta \bar W_0 (\eta_s) = \beta \bar W(x=x_0(\eta_s),\eta_s) = 
\frac{2 c^3-9 b c d + 27 a d^2 + 2(c^2 - 3 b d)^{3/2}}{27 d^2}.
\end{equation}

Beyond the position of the first maximum of the depletion potential the 
parametrized form is continued by imposing the known asymptotic behavior for
large $h$. The asymptotic behaviors of the wall-sphere and sphere-sphere 
depletion potential are slightly different and must be considered separately. 
For the wall-sphere depletion potential the asymptotic behavior is given by 
Eq.~(\ref{asymptfitw}) and the amplitude $A_w$ and the phase $\Theta_w$ are 
chosen such that the function and its first derivative are continuous at $x_0$.
From the requirement of a continuous derivative at the first maximum, i.e.,
$\left. \frac{{\mathrm d} \beta \bar W_{asympt}^w(x,\eta_s)}{{\mathrm d} x} 
\right|_{x=x_0} = 0$, the phase can be determined to be
\begin{equation} \label{Theta}
\Theta_w(\eta_s) = a_1 \sigma_s x_0 + \arccos \left( \frac{a_1}
{\sqrt{a_0^2+a_1^2}} \right).
\end{equation}
From the requirement that the function is continuous at $x_0$, i.e., 
$\bar W_{asympt}^w(x_0,\eta_s) = \bar W_0(\eta_s)$, together with the phase 
from Eq.~(\ref{Theta}), the amplitude of the asymptotic decay follows as
\begin{equation} \label{amplitude}
A_w(\eta_s) = \beta \bar W_0 \exp(a_0 \sigma_s x_0) 
\sqrt{\frac{a_0^2+a_1^2}{a_1^2}}.
\end{equation}

A similar calculation for the sphere-sphere case using Eq.~(\ref{asymptfitp})
leads to slightly different expressions for the phase,
\begin{equation} \label{Thetap}
\Theta_p(\eta_s) = a_1 \sigma_s x_0 + \arccos \left(
\frac{a_1 (x_0 \sigma_s +\sigma_b)}
{\sqrt{1+2 a_0 (x_0 \sigma_s + \sigma_b) + (a_0^2+a_1^2)
(x_0 \sigma_s +\sigma_b)^2}} \right),
\end{equation}
and the amplitude
\begin{equation} \label{amplitudep}
A_p(\eta_s)=\beta \bar W_0 \frac{\exp{(a_0 \sigma_s x_0)}}{a_1 \sigma_s} 
\sqrt{1+2 a_0 (x_0 \sigma_s + \sigma_b) + 
(a_0^2+a_1^2) (x_0 \sigma_s + \sigma_b)^2}.
\end{equation}
Unlike $\Theta_w$ and $A_w$, $\Theta_p$ and $A_p$ depend (weakly) on the size 
ratio $s=\sigma_s/\sigma_b$.

The coefficients $a$, $b$, $c$ and $d$ are fitted to depletion potentials 
calculated within DFT. Scaled depletion potentials $\bar W$ obtained for a big
hard sphere near a planar hard wall, for a size ratio $s=0.1$, are used in the
range $0\leq\eta_s\leq 0.3$. The resulting coefficients are given by
\begin{equation} \label{coeff}
\begin{array}{rcl}
a(\eta_s) & = & - 2.909~ \eta_s, \\
b(\eta_s) & = & 6.916~ \eta_s - 4.616~ \eta_s^2 + 78.856~ \eta_s^3,\\
c(\eta_s) & = & -4.512~ \eta_s + 15.860~ \eta_s^2 - 93.224~ \eta_s^3,\\
d(\eta_s) & = & -\eta_s \exp(-1.734+8.957~\eta_s+1.595~\eta_s^2).
\end{array}
\end{equation}
There is no particular significance in the chosen form of parametrization but
we note that the contact values of the scaled depletion potential 
$\beta \hat W(0,\eta_s)=a(\eta_s)$ are linear in $\eta_s$ for this choice of
parametrization. It is interesting that the coefficient $-2.909$ is rather 
close to the value $-3$ obtained from the Asakura-Oosawa result 
(valid as $\eta_s \to 0$) in the limit of small size-ratios, see 
Eqs.~(\ref{asabb}) and (\ref{asawb}) and also Eq.~(\ref{contactvalue}).

The quantities $a_0$ and $a_1$ are obtained by solving Eq.~(\ref{condition}), 
using the Percus-Yevick pair direct correlation function $c^{(2)}_{ss}(r)$. The
results are shown in Fig.~\ref{fig:a0a1}. In the range 
$0.05 \leq \eta_s \leq 0.4$ they can be fitted accurately by
\begin{equation}
a_0(\eta_s)~\sigma_s = 4.674~\exp(-3.935~\eta_s) + 3.536~\exp(-56.270~\eta_s)
\end{equation}
and
\begin{equation}
a_1(\eta_s)~\sigma_s = -0.682~\exp(-24.697~\eta_s) + 4.720 + 4.450~\eta_s.
\end{equation}
These formulae specify all the ingredients for determining the parametrized
form of the depletion potential.

\begin{figure}
\centering\epsfig{file=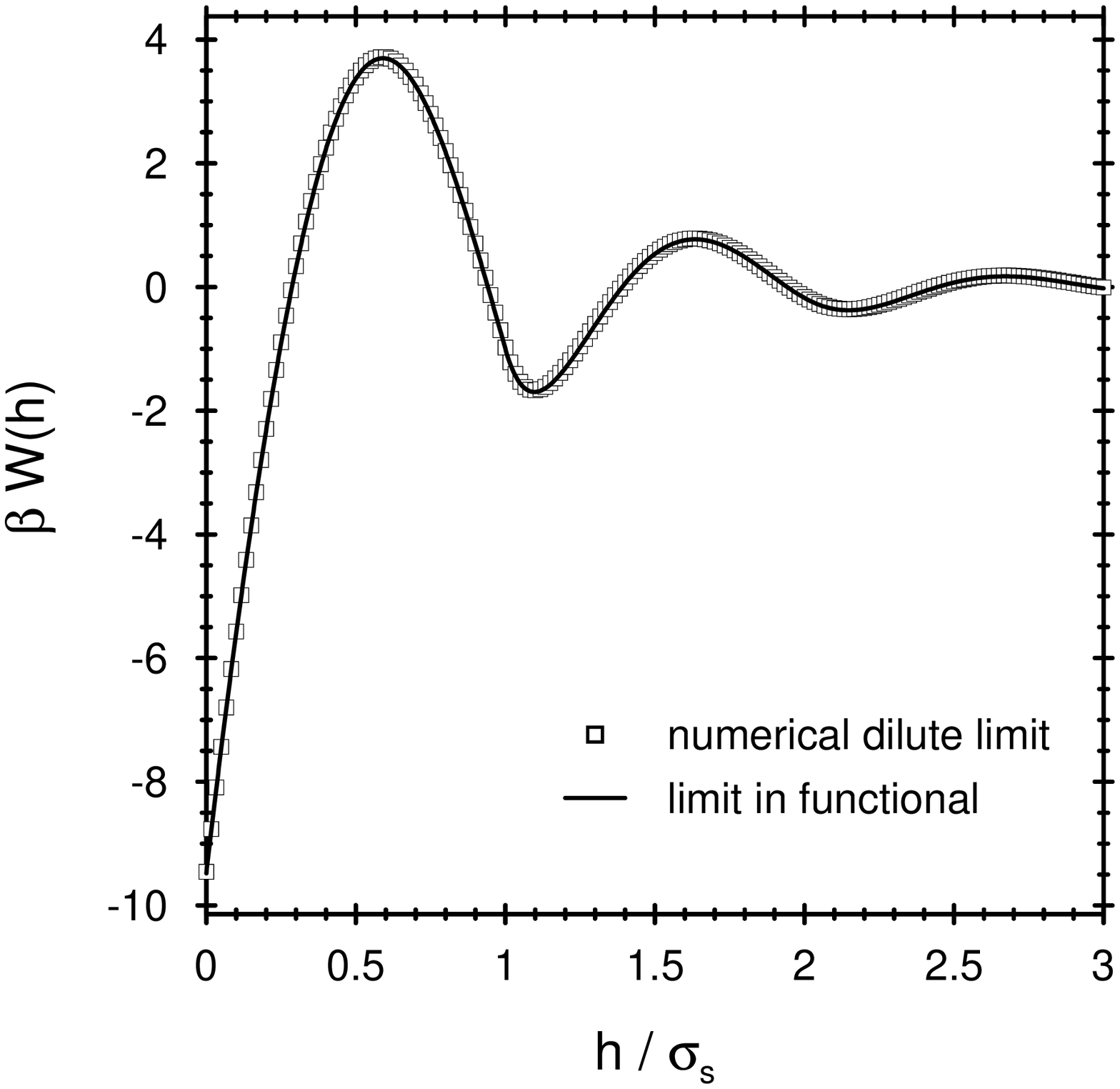,bbllx=10,bblly=60,bburx=550,bbury=580,
width=13.5cm}
\vspace{0.5cm}

\caption{\label{fig:dilute} Comparison of the depletion potential for a big 
hard sphere near a planar hard wall, calculated via two different routes. 
These correspond to taking the dilute limit directly in the functional (solid 
line), and to taking the dilute limit numerically with a packing fraction of 
the big spheres of $\eta_b=10^{-4}$ ($\square$). In both cases the size ratio 
is $s=R_s/R_b=0.1$ and the packing fraction of the small hard spheres is 
$\eta_s=0.3$. $h$ is the separation between the wall and the surface of the 
big sphere; $\sigma_s=2 R_s$ is the diameter of the small spheres.}
\end{figure}

\begin{figure}
\centering\epsfig{file=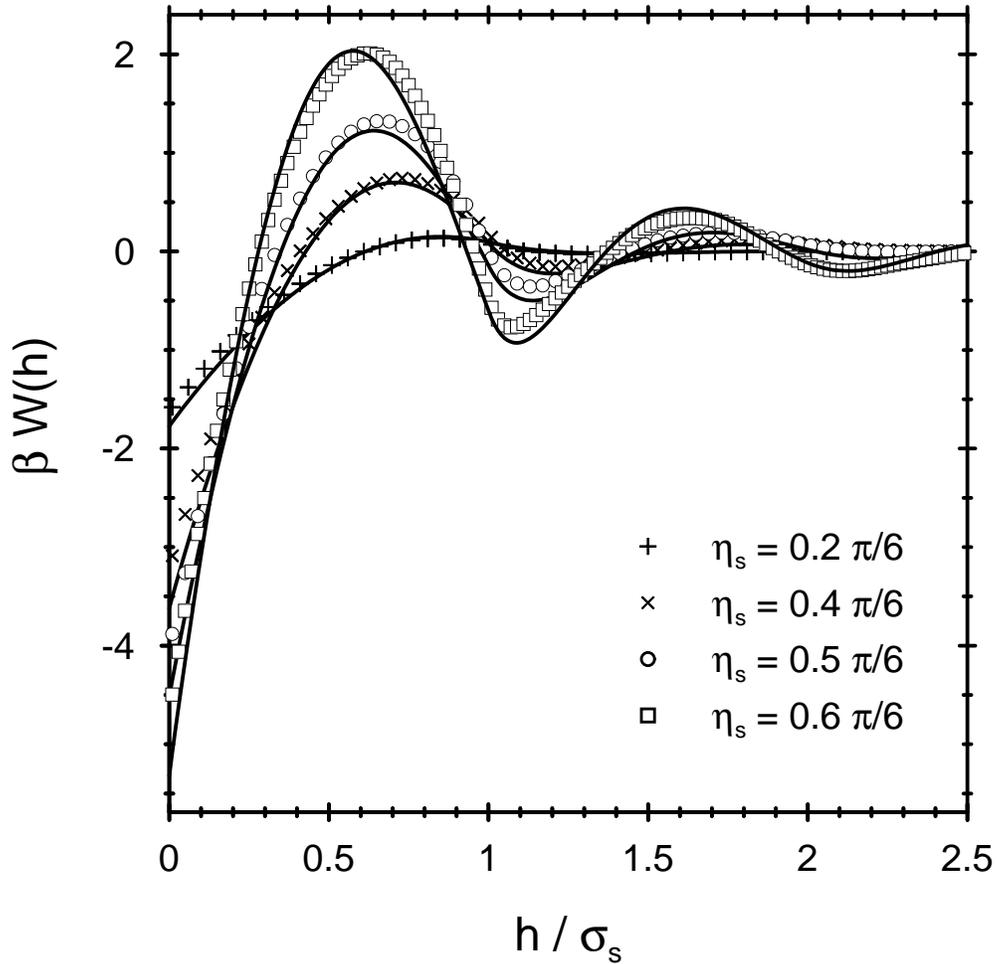,bbllx=10,bblly=60,bburx=550,bbury=580,
width=13.5cm}
\vspace{0.5cm}

\caption{\label{fig:biben} The depletion potential between two big hard spheres
in a sea of small hard spheres calculated for various packing fractions
$\eta_s$ of the small spheres and a fixed size ratio $s=0.1$. We compare 
simulation results (symbols) from Ref.~\protect\cite{Biben96} with results 
from our DFT approach (solid lines). $h$ is the separation between the 
surfaces of the two big spheres and $\sigma_s$ is the diameter of the small 
spheres.}
\end{figure}

\begin{figure}
\centering\epsfig{file=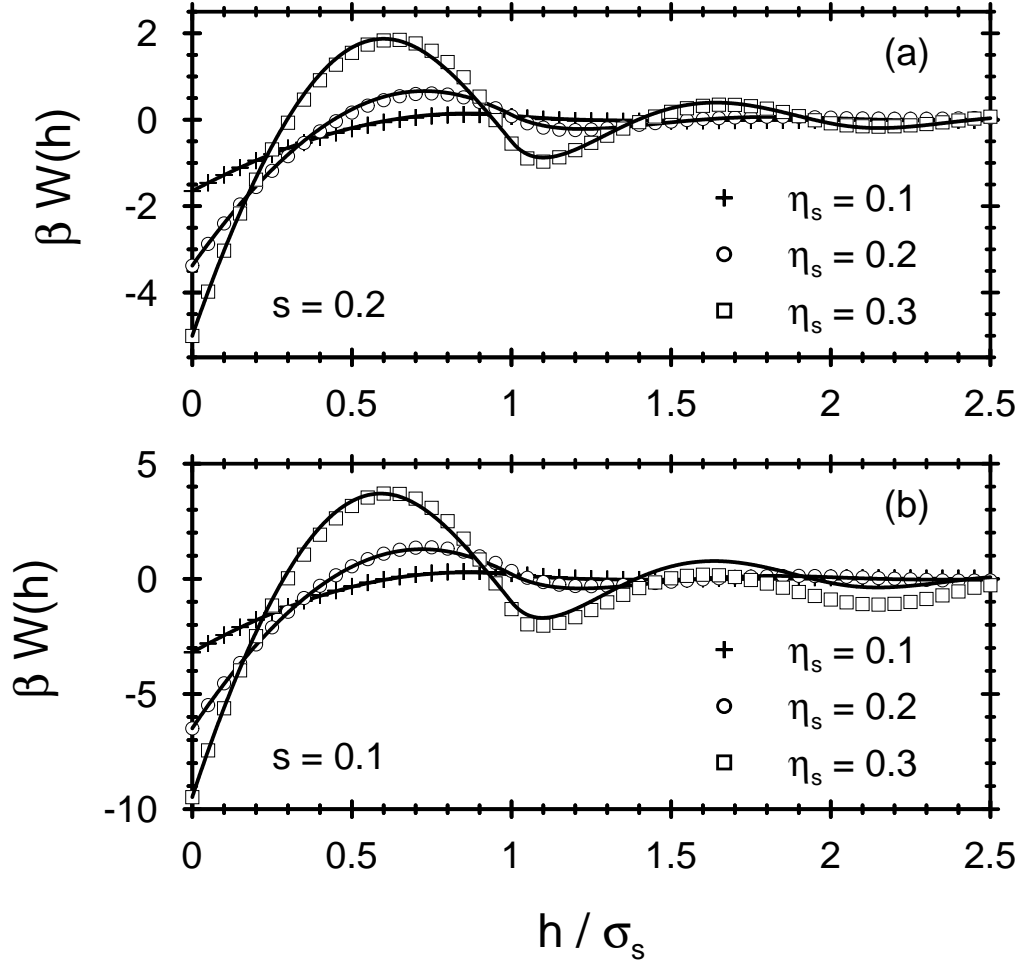,bbllx=10,bblly=60,bburx=550,bbury=580,
width=13.5cm}
\vspace{0.5cm}

\caption{\label{fig:dickman} The depletion potential between a big hard sphere 
and a planar hard wall calculated for various packing fractions $\eta_s$ of 
the small hard spheres  and size ratios $s=0.2$ (a) and $s=0.1$ (b). We 
compare processed simulation data from Ref.~\protect\cite{Dickman97} (symbols)
with those of our calculations (solid lines). The only significant deviations 
occur in (b) for $\eta_s=0.3$ (see text). $h$ is the separation between the 
wall and the surface of the big sphere; $\sigma_s$ is the diameter of the 
small spheres.}
\end{figure}

\begin{figure}
\centering\epsfig{file=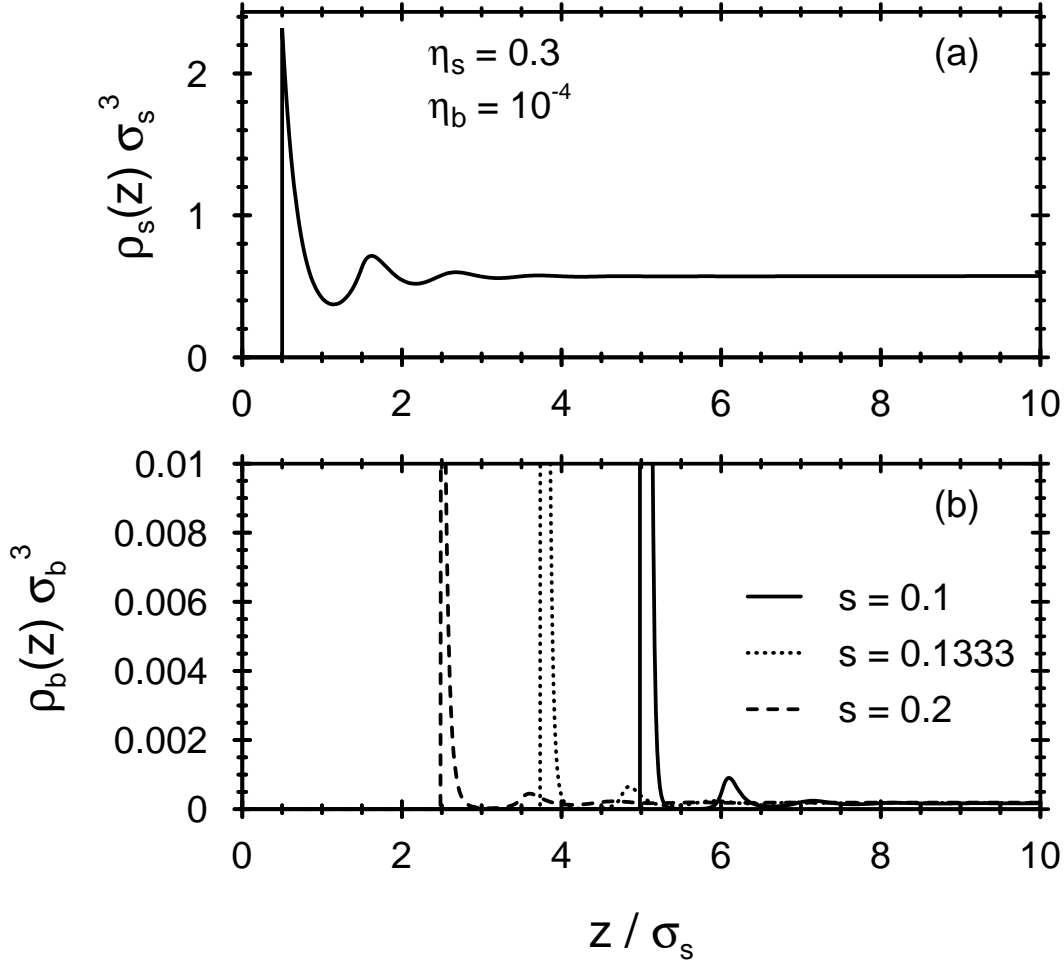,bbllx=10,bblly=60,bburx=550,bbury=580,
width=13.5cm}
\vspace{0.5cm}

\caption{\label{fig:profiles} The number density profiles of the two components
of a binary hard-sphere mixture near a planar hard wall as obtained from DFT. 
The reservoir packing fraction of the small spheres is $\eta_s=0.3$ and that of
the big spheres $\eta_b=10^{-4}$. For the three different size ratios $s=0.1$,
0.1333, and 0.2 the profiles of the small spheres (a) are indistinguishable 
while the profiles of the big spheres (b) differ considerably. The contact 
values of the density profiles of the big spheres are 
$\rho_b(z=\sigma_b^+) \sigma_b^3$ = 0.0279, 0.2518, and 2.1814 for $s$=0.2, 
0.1333 and 0.1, respectively. $z=0$ denotes the position of the wall and 
$\sigma_s$ is the diameter of the small spheres.}
\end{figure}

\begin{figure}
\centering\epsfig{file=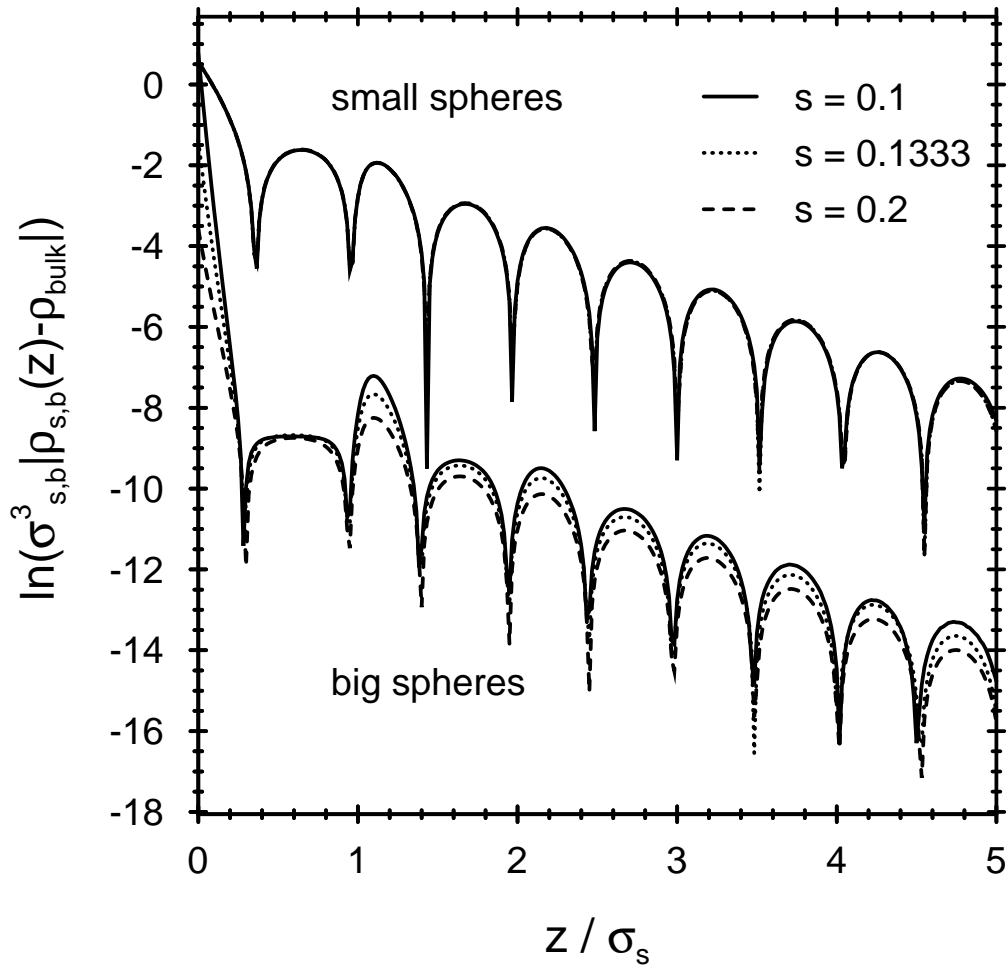,bbllx=10,bblly=60,bburx=550,bbury=580,
width=13.5cm}
\vspace{0.5cm}

\caption{\label{fig:asympt1} The asymptotic decay of the number density 
profiles of the two components of a binary hard-sphere mixture near a planar 
hard wall for three different size ratios $s$. The packing fraction of the 
small spheres is $\eta_s=0.3$ and that of the big spheres is $\eta_b=10^{-4}$. 
In each case the natural logarithm of the modulus of the density profile minus
the corresponding bulk density is plotted. The upper lines denote the density 
profiles of the small spheres; these are practically the same for all $s$. The
lower lines denote the density profiles of the big spheres and it can be seen 
that the amplitude of the oscillations does depend on the size ratio $s$. 
However, for $z/\sigma_s \gtrsim 2$ the same characteristic decay length 
$a_0^{-1}$ and wavelength of oscillation $2 \pi/a_1$ characterize the decay of 
{\em both} density profiles; i.e., for the big and the small spheres (see 
text). Note that the density profiles in Fig.~\ref{fig:profiles} have been 
shifted so that here $z$ measures the distance from contact. $\sigma_s$ is the
diameter of the small spheres.}
\end{figure}

\begin{figure}
\centering\epsfig{file=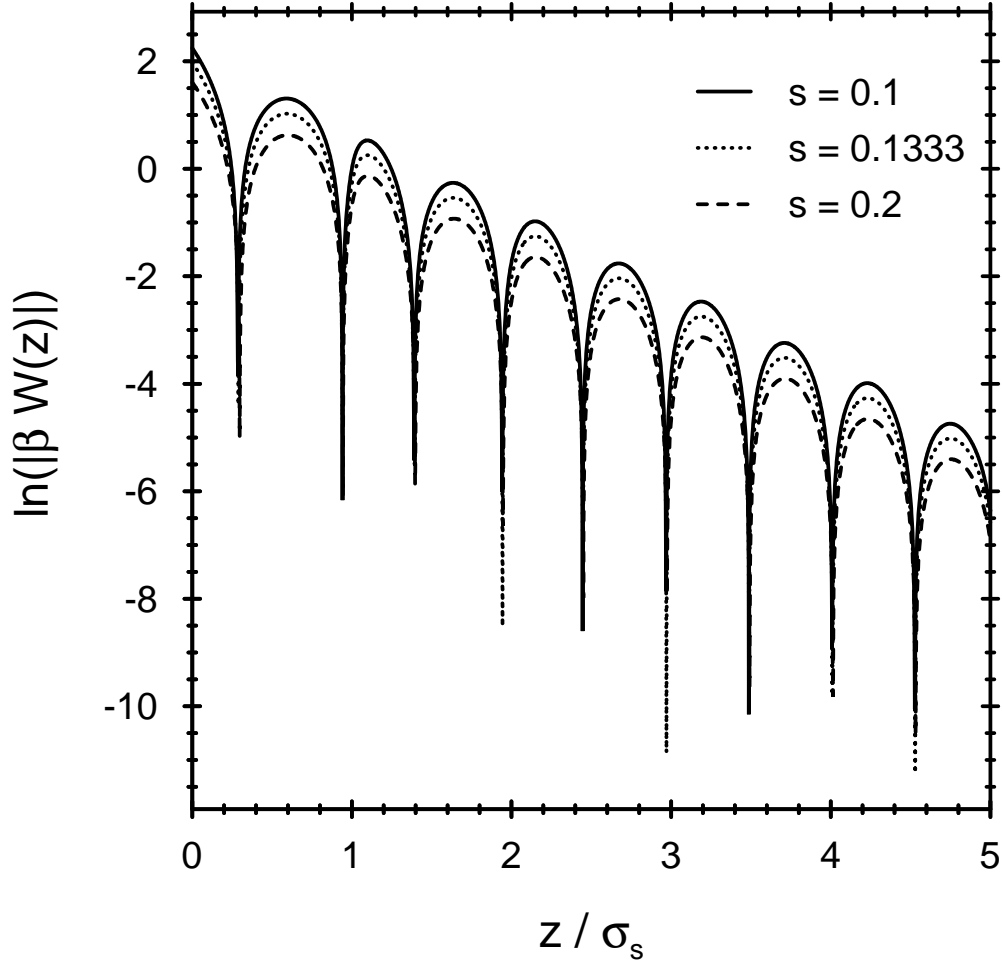,bbllx=10,bblly=60,bburx=550,bbury=580,
width=13.5cm}
\vspace{0.5cm}

\caption{\label{fig:asympt2} The asymptotic decay of the depletion potentials 
$W(z)$ for a big hard sphere near a planar hard wall for the same parameters as
shown in Fig.~\protect\ref{fig:asympt1}, i.e., $\eta_s=0.3$ and three size 
ratios $s$. The same decay length $a_0^{-1}$ and wavelength $2 \pi/a_1$ that 
determine the asymptotic decay of the density profiles determine the decay of 
$W(z)$ (see text). Only the amplitude of the oscillations depend on $s$. In 
each case $z$ measures the distance from contact. $\sigma_s$ is the diameter of
the small spheres.}
\end{figure}

\begin{figure}[htb]
\centering\epsfig{file=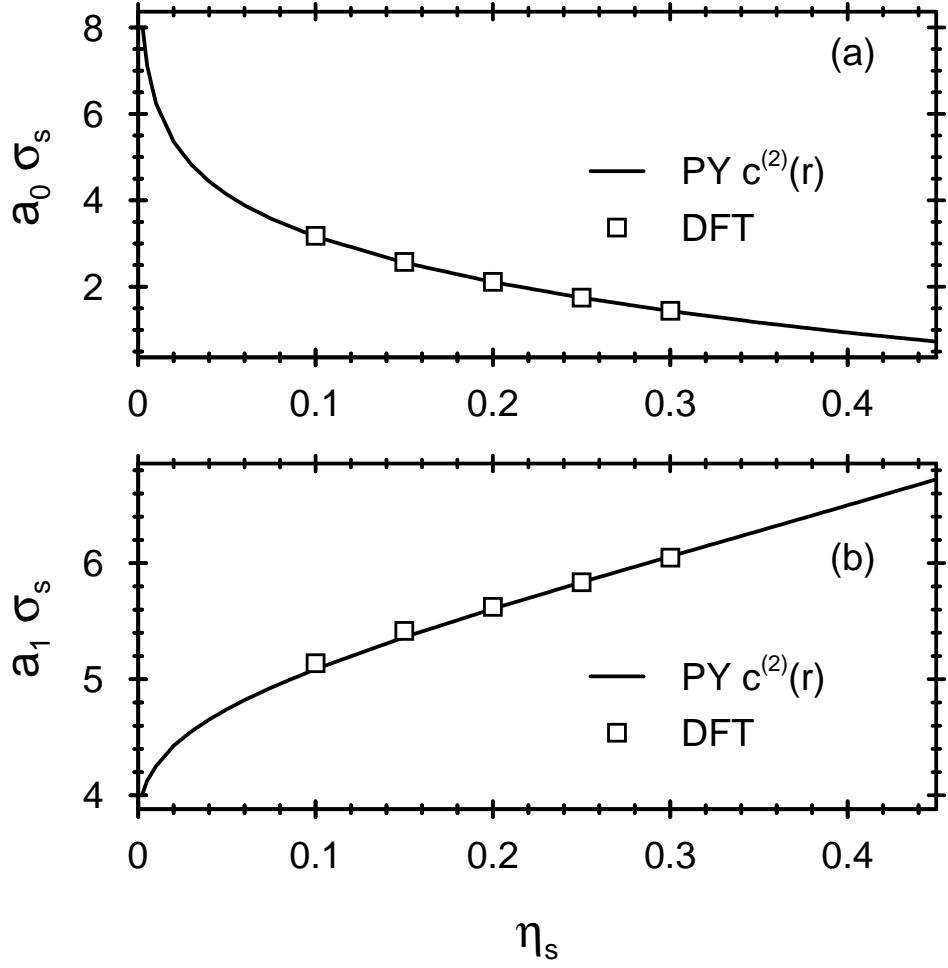,bbllx=10,bblly=60,bburx=550,bbury=580,
width=13.5cm}
\vspace{0.5cm}

\caption{\label{fig:a0a1} Comparison of (a) the inverse decay length $a_0$ and 
(b) the inverse wavelength $(2 \pi/a_1)^{-1}$ as determined from the theory of
asymptotic decay of the bulk pairwise correlation [see Eq.~(\ref{condition})]
using the Percus-Yevick two body direct correlation function $c^{(2)}_{ss}(r)$ 
(solid line) with the corresponding results obtained from density profiles 
calculated using the Rosenfeld functional ($\square$) for hard-sphere fluids 
near a planar hard wall. $\eta_s$ is the packing fraction and $\sigma_s$ is 
the diameter of the small spheres.}
\end{figure}

\begin{figure}
\centering\epsfig{file=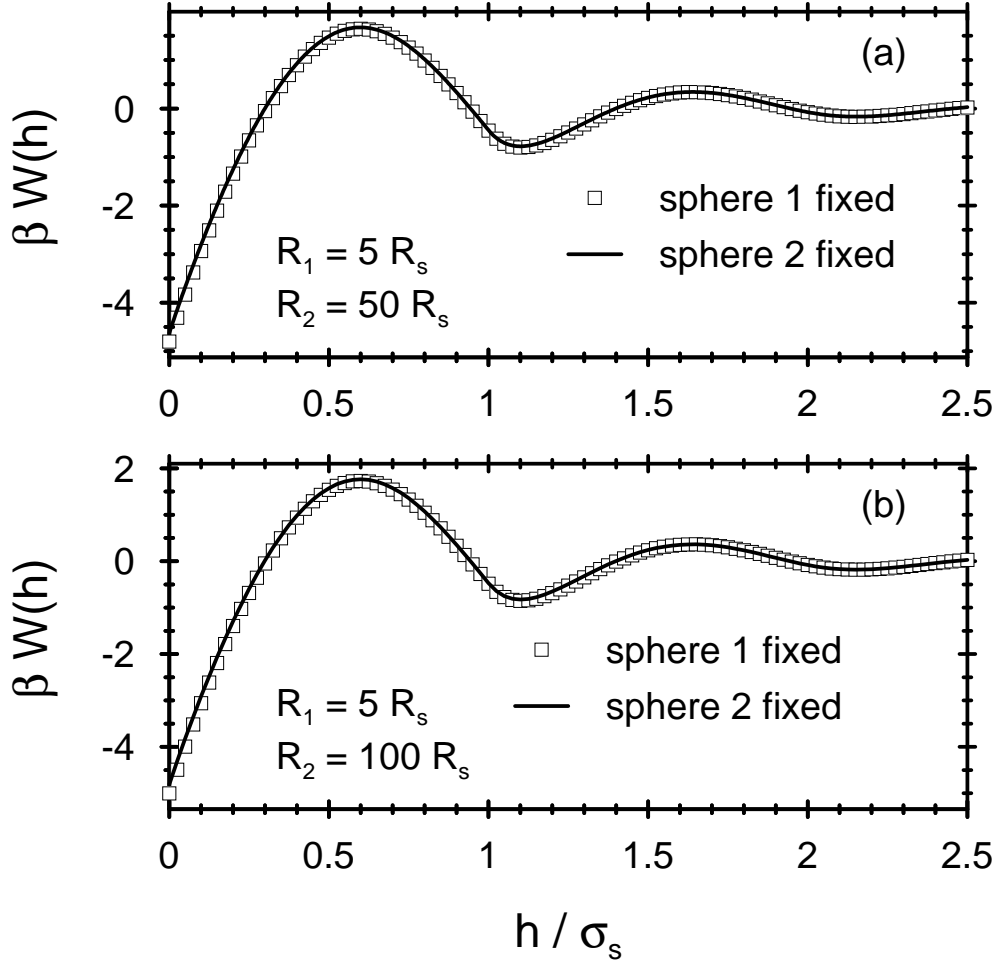,bbllx=10,bblly=60,bburx=550,bbury=580,
width=13.5cm}
\vspace{0.5cm}

\caption{\label{fig:test} The depletion potential between a hard sphere $1$ 
with radius $R_1=5 R_s$ and a hard sphere $2$ with radius $R_2=50 R_s$ (a) and 
$R_2=100 R_s$ (b), in a sea of small hard spheres with radius $R_s$ and 
packing fraction $\eta_s=0.3$. The solid line denotes the depletion potential 
calculated by fixing sphere 2 first, so that 2 enters into the calculation
as an external potential, whereas the symbols denote the depletion potential 
obtained with sphere 1 acting as the external potential. The two routes should
lead to the same results (see text). $h$ is the separation between the 
surfaces of the spheres 1 and 2 and $\sigma_s=2 R_s$.}
\end{figure}

\begin{figure}
\centering\epsfig{file=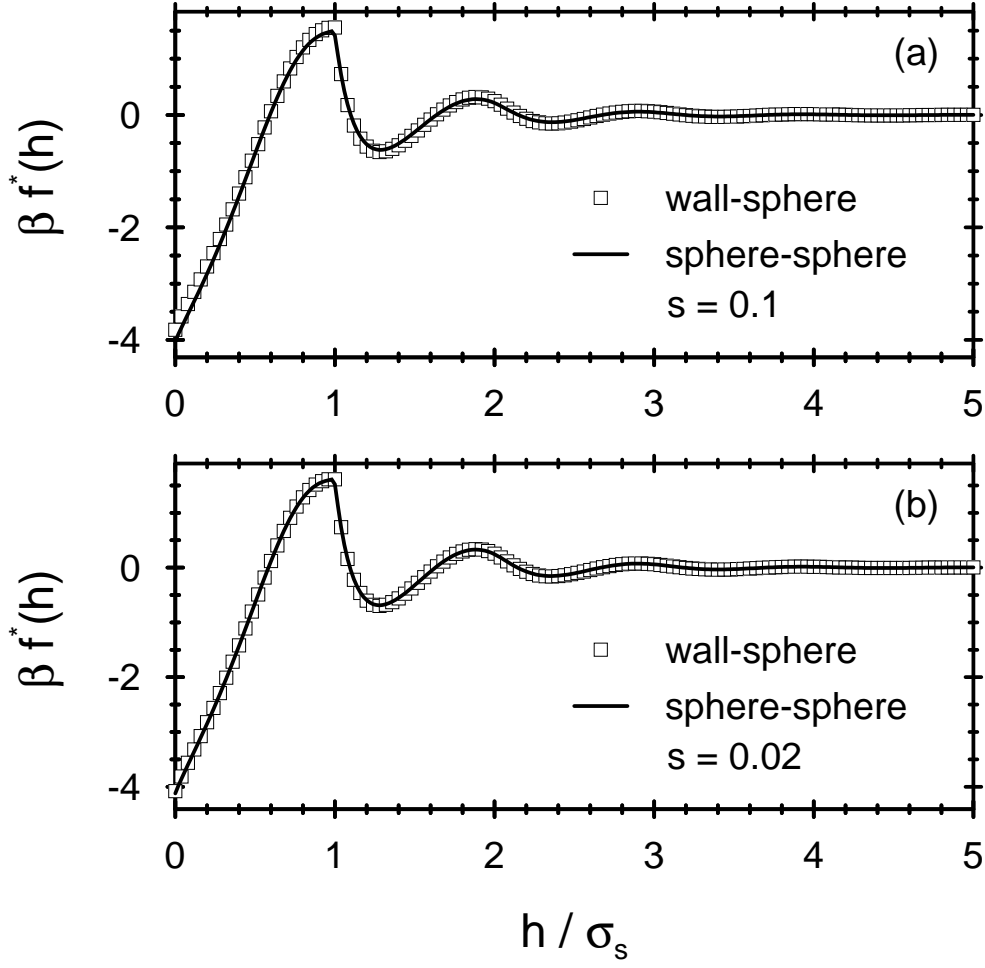,bbllx=10,bblly=60,bburx=550,bbury=580,
width=13.5cm}
\vspace{0.5cm}

\caption{\label{fig:scaling} The scaled depletion force between two big 
hard spheres $\beta f^*(h)$ = $2 \beta F(h) R_s^2/(R_b+R_s)$ (solid line) and 
between a single big hard sphere and a planar hard wall 
$\beta f^*(h)$ = $\beta F(h) R_s^2/(R_b+R_s)$ ($\square$) in a sea of small 
hard spheres at a packing fraction $\eta_s=0.3$ for size ratios $s=0.1$ (a) 
and $s=0.02$ (b). For $s=0.02$, the scaling relation in Eq.~(\ref{scaling}) is 
obeyed almost perfectly. $h$ is the separation between the surfaces of the big
spheres or between the wall and the surface of the big sphere. 
$\sigma_s = 2 R_s$ is the diameter of the small spheres.}
\end{figure}

\begin{figure}
\centering\epsfig{file=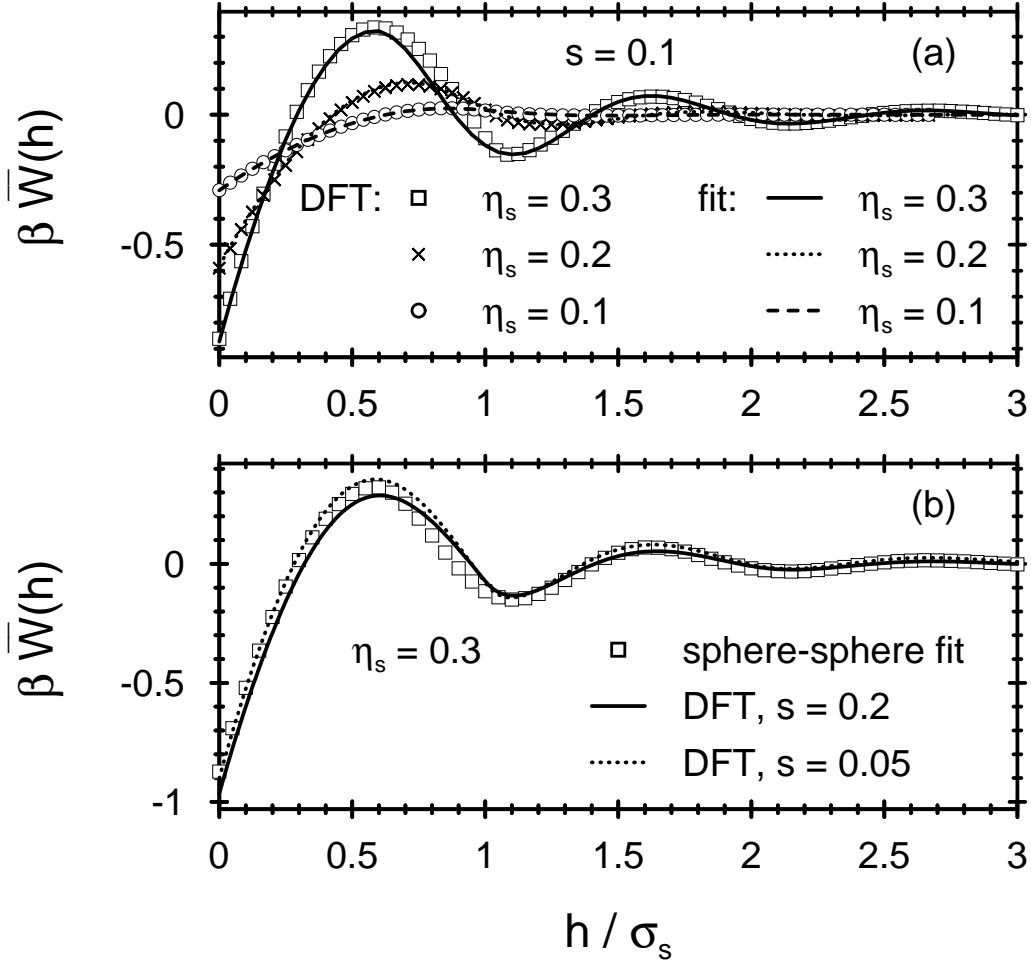,bbllx=10,bblly=60,bburx=550,bbury=580,
width=13.5cm}
\vspace{0.5cm}

\caption{\label{fig:fit} (a) Comparison of the scaled wall-sphere depletion 
potential [Eq.~(\ref{scaldep})] for various packing fractions $\eta_s$ of the 
small hard spheres and size ratio $s=0.1$ as calculated fully within DFT 
(symbols) and as given by the parametrization of Eq.~(\ref{depfit}) (lines). 
(b) Comparison of the scaled sphere-sphere depletion potentials for a packing 
fraction $\eta_s=0.3$ and size ratios $s=0.2$ and $s=0.05$ as calculated within
DFT (lines) and as given by the parametrization of Eq.~(\ref{depfit}) 
($\square$). Differences between the parametrized results for $s=0.2$ and
$s=0.05$ are not visible. $h$ is the separation between the wall and the 
surface of the big sphere or between the surfaces of the two big spheres. 
$\sigma_s$ is the diameter of the small spheres.}

\end{figure}

\begin{figure}
\centering\epsfig{file=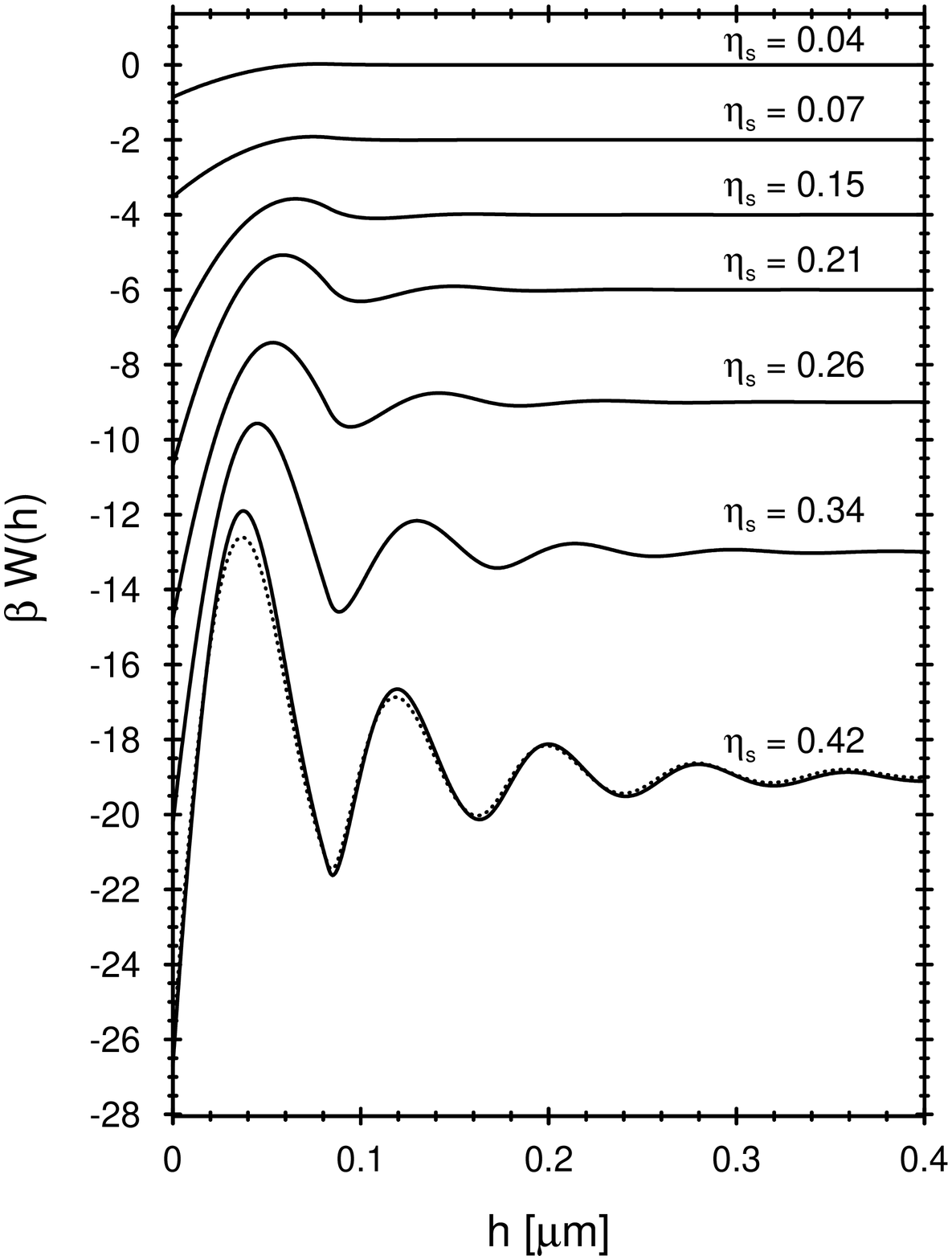,bbllx=10,bblly=60,bburx=550,bbury=770,
width=12cm}
\vspace{0.5cm}
\caption{\label{fig:crocker} The depletion potential between two big hard 
spheres in a sea of small hard spheres at various values of the 
small sphere packing fraction $\eta_s$ as obtained from the original Rosenfeld 
functional. In order to mimic the experiment of Ref.~\protect\cite{Crocker99} 
the diameters were chosen to be $\sigma_b=1.1 \mu$m and $\sigma_s=0.083 \mu$m
so that the size ratio is $s=0.0755$. $h$ measures the separation between the 
surfaces of the big spheres. Note that for display purposes each curve has been
shifted downward by a constant amount; $W(h)$ oscillates around zero as 
$h \to \infty$. The dotted line for $\eta_s=0.42$ corresponds to the depletion 
potential calculated with the modified interpolation form of the Rosenfeld 
functional \protect\cite{comment}.}
\end{figure}


\begin{references}
\bibitem{comment} Various modifications of the original Rosenfeld functional 
are proposed in Ref.~\cite{Rosenfeld96}. While all the modified functionals 
predict the freezing transition of the pure hard-sphere fluid, we employ
those that have the same bulk fluid properties as the original functional. 
In particular the functionals we choose correspond to the same (Percus-Yevick)
free-energy density and pair direct correlation functions. This implies that 
the inverse decay length $a_0$ and the wavelength $2 \pi/a_1$ of the 
asymptotic decay of the depletion potential [see Eqs.~(\ref{asymptpot1}) and 
(\ref{asymptpot2})] are the same for each choice of the functional. In our 
calculations we used the antisymmetrized version with $q=3$ [see Eq.~(27) in 
Ref.~\cite{Rosenfeld96}] and the interpolation form [Eq.~(28) in 
Ref.~\cite{Rosenfeld96}] which interpolates between the original Rosenfeld 
$\Phi$ and a form which is exact in the zero-dimensional limit.
\bibitem{comment2} The repulsive contributions reported in 
Ref.~\cite{Bechinger99} are associated with dispersion forces.
\bibitem{comment3} We suspect that several integral-equation studies of
hard-sphere mixtures have focused on states lying actually in the two-phase 
region.
\end{references}
\end{document}